\documentclass[11pt]{article}

\usepackage{amsmath,amsthm,amssymb,stmaryrd,bigints,mathabx}
\usepackage[pdftex]{graphicx}
\usepackage[titletoc,title]{appendix}
\usepackage{datetime,color,cancel,listings,authblk}
\usepackage[hyperfootnotes=false]{hyperref}
\hypersetup{
  colorlinks = true,
  linkcolor = black,
  citecolor = blue,
  urlcolor = black
}
\usepackage[margin=2.0cm]{geometry}

\usepackage[pagewise]{lineno}

\newcommand{\R}{\mathbb{R}}
\newcommand{\N}{\mathbb{N}}
\newcommand{\C}{\mathbb{C}}
\newcommand{\beq}{\begin{equation}}
\newcommand{\ee}{\end{equation}}
\newcommand{\bac}{\begin{array}{c}}
\newcommand{\ea}{\end{array}}
\newcommand{\bal}{\begin{aligned}}
\newcommand{\eal}{\end{aligned}}
\newcommand{\real}{\operatorname{Re}}
\newcommand{\imag}{\operatorname{Im}}
\newcommand{\Expe}{\operatorname{E}}

\begin{document}
\newtheorem{theorem}{Theorem}[section]
\newtheorem{lemma}[theorem]{Lemma}
\newtheorem{remark}[theorem]{Remark}
\newtheorem{observation}[theorem]{Observation}
\newtheorem{definition}[theorem]{Definition}
\newtheorem{corollary}[theorem]{Corollary}

\title{Strong solutions for the Alber equation and\\ stability of unidirectional wave spectra}

\author[1]{Agissilaos G. Athanassoulis}
\author[2]{Gerassimos A. Athanassoulis}
\author[3]{Mariya Ptashnyk}
\author[4]{Themistoklis  Sapsis}

\affil[1]{Department of Mathematics, University of Dundee}
\affil[2]{School of Naval Architecture and Marine Engineering, National Technical University of Athens}
\affil[3]{School of Mathematical and Computer Sciences, Heriot-Watt University}
\affil[4]{Department of Mechanical Engineering, Massachusetts Institute of Technology}

\maketitle

\begin{abstract}
The Alber equation is a moment equation for the nonlinear Schr\"odinger equation, formally used in ocean engineering to investigate the stability of stationary and homogeneous sea states in terms of their power spectra.  In this work we present the first  well-posedness theory for the Alber equation with the help of an appropriate equivalent reformulation.  Moreover, we show linear Landau damping in the sense that, under a stability condition on the homogeneous background, any inhomogeneities disperse and decay in time. The proof exploits novel $L^2$ space-time estimates to control the inhomogeneity and our result applies to any regular initial data (without a mean-zero restriction). Finally, the sufficient condition for stability is resolved,  and the physical implications for ocean waves are discussed. Using a standard reference dataset (the ``North Atlantic Scatter Diagram'') it is found that the vast majority of sea states  are stable, but modulationally unstable sea states do appear, with likelihood $O(1/1000);$ these would be the prime breeding ground for rogue waves.
\end{abstract}

\tableofcontents

\section{Introduction}

The Alber equation  
\beq \label{eq:Alber1}
\begin{array}{l}
\partial_t w +  2\pi p  k  \partial_x w -\\
\qquad \qquad 
-  qi\bigintsss\limits_{\lambda,y\in \R^d}  {  {e^{-2\pi i \lambda y} \Big[n(x-\frac{y}2,t)-n(x+\frac{y}2,t)\Big] dy} \,  \left(P(k-\lambda)+\epsilon w(x,k-\lambda,t)\right) d\lambda} =0,\\
\qquad \qquad  n(x,t) =  \int_{\xi\in \R^d} w(x,\xi,t) d\xi,  \qquad w(x,k,0)=w_{0}(x,k),
\ea
\ee
see e.g.\  \cite{Alber1978,alber1978stability,Athanassoulis2017,Gramstad2017,Onorato2003,Ribal2013,Ribal2013a,Stiassnie2008,Yuen1982}, {  is a  second moment of the cubic NLS equation
\beq \label{eq:NLS}
i\partial_t u + \frac{p}2 \Delta u + \frac{q}2 |u|^{2} u=0
\ee
governing the complex envelope of ocean waves, $u(x,t)$. It is derived by taking the stochastic second moment of \eqref{eq:NLS} and then using a Gaussian moment closure. Then, passing to   Wigner transform coordinates, the initial data is assumed to be close to a stationary and homogeneous background solution $P(k)$,
\[
W(x,k,t)=P(k)+\epsilon w(x,k,t), \qquad \epsilon \ll 1.
\]
Observe that the unknown $w(x,k,t)$ in equation \eqref{eq:Alber1} is the inhomogeneity.

The heuristic derivation of the Alber equation \eqref{eq:Alber1} from the NLS equation \eqref{eq:NLS} is well-known, but for completeness it is outlined in Appendix \ref{sec:app__A}. 
The Gaussian closure means this is not an exact equation for the second stochastic moment. However, in the ocean waves context  $q=o(1),$ $1/p=o(1),$ i.e. the problem is inherently weakly non-linear. This is a factor behind the empirical fact that the Gaussian closure is a meaningful one in this context \cite{Komen1994,Ochi1998}. 

The power spectrum
$P(k)$ represents  the distribution of wave energy over wavenumbers in a homogeneous sea state. Typically one would expect that  inhomogeneities  disperse, thus preserving the leading-order stationary and homogeneous character of the wavefield; indeed, this is what we find for the vast majority of plausible sea states  in Section \ref{sec:conclusions}. This would be the ``Landau damping'' / stable regime. However, in those exceptional cases where the inhomogeneity $w(x,k,t)$ is allowed to feed on the (infinite) energy of the power spectrum and grow significantly, then localized extreme events such as Rogue Waves become possible \cite{Athanassoulis2017,Bitner-Gregersen2015,ThemistoklisSapsis2016,Dematteis2017,Gramstad2017,Gramstad2018,Onorato2003,Onorato2013,Ribal2013a}. This would be the ``modulation instability'' (MI) / unstable regime, and it can be thought of as a generalisation of the standard MI of the NLS \cite{Benjamin1967,zakharov1968stability,Zakharov2009} to continuous spectra. 

The criterion for (in)stability involves only the power spectrum $P(k),$ and is related to the  ``eigenvalue relation'' which appeared in  \cite{Alber1978} as a sufficient condition for instability. Some refinements are required, and the relation between the different kinds of (in)stability conditions is the object of Theorem \ref{thrm:PenroseConditionResolve} (see also Remark \ref{rm:vgcfderr4543}). A key fact here is that the bifurcation from Landau damping to MI involves only the shape of the power spectrum $P(k),$ and is not sensitive to the initial inhomogeneity.  Determining whether a power spectrum is stable or unstable is a crucial question in the oceanographic context, and presents certain challenges \cite{Gramstad2017,Ribal2013a,Stiassnie2008}. This is discussed in some detail in Section \ref{sec:conclusions}, where Theorem \ref{thrm:PenroseConditionResolve} is used in a novel, straightforward way to check the stability of a given spectrum $P(k)$. 

While the Alber equation has been used formally for some time,  there are still many open questions related to it. It has only been  recently that
 some works for well-posedness and stability of related nonlinear  equations have appeared. 
  In  \cite{Lewin2014, Lewin2015} the authors work in operator formalism, for a defocusing problem $(p\cdot q<0)$ with  a regular interaction kernel\footnote{That is,  $K(x)\ast n(x,t)$ appears in the equation instead of $n(x,t).$ Whenever $n(x,t)$ appears directly, as in equation \eqref{eq:Alber1}, this is also called a $\delta$ interaction kernel, since $n(x,t)=\delta(x)\ast n(x,t).$}. The authors  exploited the defocusing character of the problem by defining a relative entropy which controls the solution in an appropriate sense; this is a key ingredient of their proof.  In \cite{Chen2015}  a similar  argument is used for the defocusing problem  with a $\delta$ interaction kernel and with a single background spectrum. Another related work is \cite{DeSuzzoni2015}, where the stability of a fully stochastic problem (no Gaussian closure) is studied, but   only in the defocusing case,  $d\geqslant 4$ and with a smooth interaction kernel. 
 
 More broadly, there are analogies between the classical Landau damping problem  for the Vlasov equation \cite{Mouhot2011,Penrose1960} and the stability of the Alber equation. The most closely related work from that context seems to be \cite{Bedrossian2018} where
 the Vlasov equation is studied in $d=3$ and with mean-zero initial data, as opposed to $d=1$ (leading to weaker dispersion) and general initial data, which is the natural setting for the Alber equation.}

{ This paper is organised as follows: in Section~\ref{Section_2}  some definitions and notations are summarized. The main results are formulated in Section~\ref{sec:mainresults}.  We show well-posedness and regularity of solutions for any dimension in Theorems \ref{thrm:wfi_lwp}, \ref{cor:RegSol}, \ref{thrm:lwfi} below. In Theorem \ref{thrm:MainTheorem} we consider the one-dimensional case and  show for the first time that  the  inhomogeneities decay in time under a stability condition, using   
 novel $L^2$ space-time estimates (cf. Lemmata \ref{lm:freespace1} and \ref{lm:ntonfforsomeindices}).}
In Theorem \ref{thrm:PenroseConditionResolve} we derive the stability condition, and show it is complementary to the sufficient condition for instability, the``eigenvalue relation'' mentioned earlier.  Using Theorem \ref{thrm:PenroseConditionResolve} we investigate the stability of sea states in the North Atlantic  in Section~\ref{sec:conclusions}. {The proofs of main results are given in Sections~\ref{sec:proofsexistence1}-\ref{sec:PenrCond765} and in Section~\ref{sec:conclusions}  we  discuss  applications.}

\section{Mathematical preliminaries}\label{Section_2}
{We shall start with summarising main notations and definitions used in the statement and proofs of main results. }

\subsection{Definitions and notations}\label{subsec:defs_notations}

The normalisation we use for the Fourier transform is
\[
\widehat{u}(X)=\mathcal{F}_{x\to X}[u] = \int\limits_{x\in\R^d} e^{-2\pi i x\cdot X} u(x)dx, \qquad 
\widecheck{u}(X)=\mathcal{F}^{-1}_{x\to X}[u] = \int\limits_{x\in\R^d} e^{2\pi i x\cdot X} u(x)dx
\]
{for $X \in \R^d$. }

\begin{definition}[Spaces of bounded derivatives and moments] \label{def:spaces}
Consider a function on phase-space $f(x,k),$ $s\in \N,$ $p\in[1,\infty].$ 
The $\Sigma^{s,p}$ norm will be defined as
\[
\|f\|_{\Sigma^{s,p}} = \sum\limits_{0\leqslant|a+b+c+d|\leqslant s} \|x^ak^b\partial_x^c\partial_k^d f\|_{L^p(\R^{2d})}.
\]
We will also use the
standard Sobolev spaces
$
\|f\|_{W^{s,p}} = \sum\limits_{0\leqslant|c+d|\leqslant s} \|\partial_x^c\partial_k^d f\|_{L^p},$ $\|f\|_{H^{s}}=\|f\|_{W^{s,2}}.
$
\end{definition}

One readily checks the following
\begin{lemma}[Embeddings of the $\Sigma^{s,p}$]\label{lm:Sigmasebeddings}
By virtue of the Sobolev embeddings, 
\beq\label{eq:sobemb}
\forall q,p\in[1,\infty], \,\, s\in\N \qquad \exists s_0\in\mathbb{N}, \,\, C>0 \qquad \|f\|_{\Sigma^{s,p}} \leqslant C \|f\|_{\Sigma^{s+s_0,q}}.
\ee
Denoting $\mathcal{S}(\R^{2d})$ the Schwarz class of test-functions on phase-space, observe that for any $q\in[1,\infty]$
\[ 
\bigcap_{s\in \N} \Sigma^{s,q}(\R^{2d}) = \mathcal{S}(\R^{2d}).
\]
Moreover, the spaces $\Sigma^{s,2}$ are closed under Fourier transforms in the sense that
\[
\mathcal{F}_{(x,k)\to (X,K)}\Sigma^{s,2}=\mathcal{F}_{k\to K}\Sigma^{s,2}=\mathcal{F}_{x\to X}\Sigma^{s,2}=\Sigma^{s,2}
\]
and similarly for inverse Fourier transforms. Combined with equation \eqref{eq:sobemb}, this means that 
\[
\forall q,p \in [1,\infty], \,\, s\in\N \qquad \exists s_0\in \N, \,\, C>0 \quad \mbox{ such that } \quad \|f\|_{\Sigma^{s,q}} \leqslant C \| \mathcal{F} f\|_{\Sigma^{s+s_0,p}}
\]	
where $\mathcal{F}$ denotes a forward or inverse Fourier transform in the $x,$ $k,$ or $(x,k)$ variables.
\end{lemma}

We will also use the Laplace transform, denoted as
$
\widetilde{u}(\omega):=\mathcal{L}_{t\to\omega}[f] = \int_{t=0}^{+\infty} e^{-\omega t} u(t) dt,
$
and the Hilbert transform $\mathbb{H}$ and the signal transform $\mathbb{S}$
\beq\label{eq:define_hilbert_signal_ops}
\mathbb{H}[u](x)=\frac{1}\pi \mbox{p.v.} \int\limits_{t\in\R} \frac{u(t)}{x-t}dt, \qquad \qquad 
\mathbb{S}[u](x)= \mathbb{H}[u](x) - i u(x),
\ee
respectively.

In the context of the inverse Laplace transform we will also  use an alternate  ``Fourier transform in time'',
\[
\mathfrak{F}_{t\to s}[u(t)] := \int\limits_{t\in\R} e^{-ist} u(t)dt, \qquad \mathfrak{F}^{-1}_{s\to t}[v(s)]=\frac{1}{2\pi} \int\limits_{s\in\R} e^{ist} v(s)ds.
\]

Obviously
\[
\|\mathfrak{F}[u]\|_{L^2}=\sqrt{2\pi}\|u\|_{L^2}, \qquad \|\mathfrak{F}^{-1}[v]\|_{L^2}=\frac{1}{\sqrt{2\pi}}\|v\|_{L^2}, \qquad \mathfrak{F}[tu(t)] = i \partial_s \mathfrak{F}[u].
\]

In the statement and proof of the main results we will also use the following
\begin{definition}[$D_XP$] For a function $P:\R\to\R$ we will use the notation
\[
D_XP(k) = \left\{ \begin{array}{cl}
\frac{P(k+\frac{X}2)-P(k-\frac{X}2)}{X}, & X\neq 0\\
P'(k), & X=0.
\end{array}
\right.
\]	
\end{definition}

By abuse of notation all constants will be denoted by $C,C',C''.$ To keep track of dependence on important parameters we will use e.g. $C=C(t,p,q).$

\subsection{Reformulation of the problem and heuristics}

To study  problem  \ref{eq:Alber1} it is helpful to use equivalent reformulations. 
If we take the inverse Fourier transform in $x$ of the original Alber equation we pass to the  Alber-Fourier equation
\beq \label{eq:linearisedF_11}
\begin{aligned} 
&\partial_t f - 4\pi^2 i p k \cdot X f 
+qi \left[ P\Big(k-\frac{X}2\Big) - P\Big(k+\frac{X}2\Big) \right] \widecheck{n}(X,t)+
\\
&+\epsilon qi \int_{\R^d} \widecheck{n}(y,t)  \left[ f\Big(X-y,k-\frac{y}2\Big) - f\Big(X-y,k+\frac{y}2\Big) \right]dy=0, 
\\ 
f(X,&k,0)=f_0(X,k)=\mathcal{F}^{-1}_{x\to X} [w_{0}], \qquad 
\widecheck{n}(X,t)=\int_{\R^d} f(X,\xi,t)d\xi=\mathcal{F}^{-1}_{x\to X}[n(x,t)], 
\end{aligned} 
\ee
{where
$$f(X,k,t) := \mathcal{F}_{x\to X}^{-1}[w(x,k,t)]=\int\limits_{\R^d} e^{2\pi i x\cdot X} w(x,k,t)dx.$$
 }

To motivate linear stability, let us start from the linearised problem,
\beq \label{eq:LinAlbMOVETOSTART}
\begin{aligned}
& \partial_t f - 4\pi^2 i p k \cdot X f
+qi \Big[ P\Big(k-\frac{X}2\Big) - P\Big(k+\frac{X}2\Big) \Big] \widecheck{n}(X,t)=0, 
\\
f(&X,k,0)=f_0(X,k)=\mathcal{F}^{-1}_{x\to X}[w_0], \qquad 
\widecheck{n}(X,t)=\int_{\R^d} f(X,\xi,t)d\xi=\mathcal{F}^{-1}_{x\to X}[n(x,t)].
\end{aligned} 
\ee
By recasting in mild form we have
\beq \label{eq:LinAlbMOVETOSTARTmild}
f(X,k,t) - e^{4\pi^2 i p k \cdot X t} f_0(X,k)= 
 -qi \int\limits_{0}^t e^{4\pi^2 i p k\cdot X (t-\tau)}  
\Big[ P\Big(k-\frac{X}2\Big) - P\Big(k+\frac{X}2\Big) \Big] \widecheck{n}(X,\tau)d\tau,
\ee
and by integrating in $k$ we obtain a closed problem for  $\widecheck{n}(X,t),$
\beq \label{eq:LinAlbMOVETOSTARTmildposden}
\begin{aligned} 
\widecheck{n}(X,t) - \widecheck{n}_f(X,t) &=  \int_{0}^t h(X,t-\tau) \widecheck{n}(X,\tau)d\tau=0, 
\qquad 
h(X,t) &=2q\, \sin(2\pi^2pX^2t) \widecheck{P}(2\pi p Xt),
\end{aligned}
\ee
where
$n_f(x,t)$ is the known ``free-space position density'',
\beq \label{eq:def_n_f}
n_f(x,t) := \int\limits_{\R^d} w_{0}(x-2\pi p kt,k) dk \quad \Rightarrow \quad  \widecheck{n}_f(X,t) = \mathcal{F}^{-1}_{x\to X}[n_f(x,t)] = \int\limits_{\R^d} e^{4\pi^2 i p k \cdot X t} f_0 \, dk.
\ee
Now denote for brevity
\beq\label{eq:defntilde}
\widetilde{n}(X,\omega):=\mathcal{L}[\widecheck{n}(X,t)], \qquad
\widetilde{n}_f(X,\omega):=\mathcal{L}[\widecheck{n}_f(X,t)], \qquad 
\widetilde{h}(X,\omega):=\mathcal{L}[h(X,t)];
\ee
by taking the Laplace transform of equation \eqref{eq:LinAlbMOVETOSTARTmildposden} and rearranging terms we obtain 
\beq\label{eq:laplastartpoint}
\bac
\widetilde{n}(X,\omega) = \widetilde{n}_f(X,\omega) + \widetilde{h}(X,\omega)\widetilde{n}(X,\omega) \quad 

\Rightarrow \quad X\widetilde{n}(X,\omega) - X\widetilde{n}_f(X,\omega) = \frac{\widetilde{h}(X,\omega)}{1-\widetilde{h}(X,\omega)} X\widetilde{n}_f(X,\omega).
\ea
\ee
This last equation will be the starting point for the proof of Theorem \ref{thrm:MainTheorem} in Section \ref{sec:proof} (where the Laplace transforms will also be justified). For now it should clearly motivate the following

\begin{definition}[Stability condition]\label{def:PenStable1}
We will say that a spectrum $P\in\mathcal{S}(\R)$  with   compact support is {stable} if there is some $\kappa>0$ such that
\beq\label{eq:The_Penrose_Stability}
\inf\limits_{\substack{\real\omega>0,\\X\in\R}} |1-\widetilde{h}(X,\omega)| \geqslant \kappa >0.
\ee
\end{definition}

\section{Main results} \label{sec:mainresults}
{Here we state the main results of the paper.} 

\begin{theorem}[Local well-posedness in $L^1$ for the Alber-Fourier equation] \label{thrm:wfi_lwp}
Let $f_0 \in L^1(\mathbb{R}^{2d}),$ $P\in L^1(\R^d).$ Then there exists a maximal time $T=T(\|f_0\|_{L^1(\R^{2d})},q,\epsilon,\|P\|_{L^1(\R^d)})>0$ such  that there exists a unique mild solution $f(t) \in C ([0,T), L^1(\mathbb{R}^{2d}))$ of equation \eqref{eq:linearisedF_11}.

Moreover, the blowup alternative holds, i.e.\ either $T=+\infty$ or $\lim\limits_{t\to T^-}\|f(t)\|_{L^1(\R^{2d})}=+\infty$.
\end{theorem}

The proof can be found in Section  \ref{subsec:existyence}. 

\begin{theorem}[Higher regularity for  solutions of the nonlinear problem]\label{cor:RegSol} Denote $f(t)$ the solution of equation \eqref{eq:linearisedF_11} with initial data $f_0\in\mathcal{S}(\R^{2d}),$ and $T=T(\|f_0\|_{L^1(\R^{2d})},q,\epsilon,\|P\|_{L^1(\R^d)})$ 	as in Theorem \ref{thrm:wfi_lwp}. Assume moreover that $P\in\mathcal{S}(\R^d).$ 
Then 
\beq\label{eq:reg00122}
f(t) \in \mathcal{S}(\R^{2d}) \qquad \forall t\in [0,T).
\ee
Moreover, 
\beq\label{eq:reg00132z}
f\in C^\infty([0,T),\Sigma^{s,1}) \qquad \forall s\in\N.
\ee
\end{theorem}

Theorem \ref{cor:RegSol} is proved in Section  \ref{subsec:propergef887}. Combined with Lemma \ref{lm:Sigmasebeddings} it yields local-in-time well-posedness and regularity of solutions for the Alber equation \eqref{eq:Alber1}.


\begin{theorem}[Global well-posedness and exponential bounds  for the linearised problem] \label{thrm:lwfi}
Denote $f(t)$ the solution of the linearised Alber-Fourier equation \eqref{eq:LinAlbMOVETOSTART} with initial data $f_0\in\mathcal{S}(\R^{2d}).$
Assume moreover  $P\in\mathcal{S}(\R^d).$
Then the maximal time is $T=+\infty$ for all initial data  and for each $s\in \mathbb{N}$ there exists some $C=C(s,d,q,P)>0$ so that
\beq\label{eq:gfxclaim1yg}
\| f(t)\|_{\Sigma^{s,1}} \leqslant \| f_0\|_{\Sigma^{s,1}} Ce^{C t}.
\ee
Moreover,
there exist some $s_2=s_2(d)$ and  $C=C(s_2,d,q,P)$  so that
\beq\label{eq:gfxclaim2yg}
\|\widecheck{n}(t)\|_{L^\infty_X}+\|\partial_t\widecheck{n}(t)\|_{L^\infty_X}+\| \partial_t f(t)\|_{L^\infty_{X,k}}  \leqslant \| f_0\|_{\Sigma^{s_2,1}} C e^{C t}.
\ee
\end{theorem}

The proof can be found in Section \ref{subsec:propergef887}.

\begin{theorem}[Linear stability for the Alber equation in $d=1$] \label{thrm:MainTheorem} Let $P\in\mathcal{S}(\R)$ be a background spectrum of compact support which is {stable}  in the sense of Definition \ref{def:PenStable1}. Consider the linearised Alber equation 
\beq \label{eq:linearisedAlh8eia}
\begin{aligned} 
 \partial_t w + 2\pi p k \cdot \partial_x w & 
- q\, i  \bigintsss_{\lambda,y\in \R} {  {e^{-2\pi i \lambda y} \Big[n\big(x+\frac{y}2,t\big)-n\big(x-\frac{y}2,t\big)\Big] dy }\,  P(k-\lambda)   }\,  d\lambda  =0,
\\ 
& n(x,t)=\int_{\xi \in \R} w(x,\xi,t)d\xi, \qquad w(x,k,0)=w_0(x,k)\in\mathcal{S}(\R^2).
\end{aligned} 
\ee
Then there exists $r\in\N$ large enough so that the force $\partial_x n(x,t)$ decays in time in the sense that
\beq\label{eq:landaydampingstatement}
\|\partial_x n\|_{L^2_{x,t}} \leqslant C { \frac{\kappa+1}{\kappa^2}} \|w_0\|_{\Sigma^{r,\infty}}.
\ee
Furthermore, denoting $E(t):w_0(x,k)\mapsto w_0(x-2\pi p k t,k)$ the free-space propagator, there exists a wave operator $\mathbb{W}$ so that
\beq\label{eq:waveoperatorstatement}
\lim\limits_{t\to \infty}\| w(t) - E(t) \mathbb{W}(w_0) \|_{L^\infty(\R^2)} =0.
\ee
\end{theorem}
The proof is  given in Section~\ref{sec:proof}.

\begin{theorem}[Equivalent formulations of the  stability condition] \label{thrm:PenroseConditionResolve}
Let $P(k)\in\mathcal{S}(\R)$ be the background spectrum. Assume moreover that $P$ is of compact support.
Then the following statements are equivalent:
\begin{enumerate}
	\item[(A).] $\inf\limits_{\substack{\real\omega>0,\\X\in\R}} |1-\widetilde{h}(X,\omega)|=0,$ i.e. the spectrum is not  stable in the sense of Definition \ref{def:PenStable1}.
\item[(B).]
\[
\exists \,\, X_*\in\R, \quad \Omega_*\in\C \setminus\R  \quad \mbox{ such that } \quad 
\mathbb{H}[D_{X_*}P](\Omega_*)=\mathbb{H}[D_{X_*}P](\overline{\Omega}_*)=\frac{4\pi p}{q}
\]
or 
\[
\exists \,\, X_*,\Omega_*\in\R \quad \mbox{ such that } \quad 
\mathbb{H}[D_{X_*}P](\Omega_*)=\frac{4\pi p}{q} \,\, \mbox{ and } \,\, D_{X_*}P(\Omega_*)=0.
\]
\item[(C).] $d(\overline\Gamma,4\pi p/q) =0,$ where 
\beq\label{eq:defcurveG}
\bac
\Gamma_X:= \left\{\mathbb{S}[D_X P(\cdot)](t), \,\, t \in \R\right\}\cup \{0\}, \quad \overset{\circ}{\Gamma}_X=\{ z\in\C |  z \mbox{ enclosed by } \Gamma_X\}, \quad  
\overline\Gamma := \overline{\bigcup\limits_{X\in\R} \overset{\circ}{\Gamma}_X}.
\ea
\ee   
\end{enumerate}

Moreover, we have the following sufficient condition for stability: if
\[
\forall t_* \mbox{ such that } D_XP(t_*)=0 \mbox{ the condition } \mathbb{H}[D_XP](t_*) < \frac{4\pi p}{q} \mbox{ holds }
\]
then $P$ is  stable in the sense of Definition \ref{def:PenStable1}.
\end{theorem}

The proof can be found in Section \ref{sec:PenrCond765}. An implementation of the criterion (C) above in the context of ocean engineering   is visualised in Figures \ref{Fig:1} and \ref{Fig:3} and discussed in Section \ref{sec:conclusions}.

\begin{remark}[Stability condition and Alber's nonlinear eigenvalue relation]\label{rm:vgcfderr4543}
In \cite{Alber1978}  a two-dimensional setup is used, but the spectrum is integrated in the transverse direction, leading to an effective one-dimensional spectrum and a condition on it.
This one-dimensional ``eigenvalue relation'' in our notation and scalings becomes
\beq\label{eq:AlpberCondinst}
\exists X_*\in\R,\,\,\, \real(\omega_*)>0 \quad \mbox{ such that } \quad 
qi
\int_\R \frac{P(k+\frac{X_*}2) - P(k-\frac{X_*}2)}{\omega_* - 4\pi^2 i p k \, X_*}dk=1.
\ee
If it is satisfied  then linear instability follows.  To see the relationship between this condition and (B) of Theorem \ref{thrm:PenroseConditionResolve} above observe that for $X_*\neq 0,$ $\Omega_*:=\omega_*/(4\pi pi X_*)$ equation \eqref{eq:AlpberCondinst} becomes
\[
\exists 0\neq X_*\in\R,\,\, \Omega_* \in \C \,\, \mbox{ with } \,\, \operatorname{sign}(X_*) \cdot \imag(\Omega_*)<0 \quad \mbox{ such that } \quad 
\mathbb{H}[D_{X_*}P](\Omega_*)=\frac{4\pi p}{q}.
\]
The form (B) in Theorem \ref{thrm:PenroseConditionResolve} appropriately takes into account the case $X_*=0$ as well (equation \eqref{eq:AlpberCondinst} by construction has no solutions for $X_*=0,$ but stability may still fail due to what could be called renormalised solutions corresponding to $X_*=0$).

\end{remark}

{ 
\begin{remark}[Compact support assumption for $P(k)$ in the main results] In Theorem \ref{thrm:MainTheorem}  the assumption that $P(k)$ has compact support is made. This allows for Theorem \ref{thrm:l1hilbertzerointegral} to be invoked in Section \ref{sec:proof} so that the integrability requirement of equation \eqref{eq:L1constraint} in Theorem \ref{thrm:InvLaplaceOPen}, itself a central ingredient of the proof, is satisfied. The same assumption is also needed for Lemma \ref{lm:decay_htilde}, which is invoked in the proof of Theorem \ref{thrm:PenroseConditionResolve}. 
So it seems that in the current version of the paper the compact support requirement cannot be removed, although this might eventually be possible with other techniques.

What does this mean in terms of the physical application in Section \ref{sec:conclusions}? Many widely used ocean power spectra involve power decay at infinity, $P(k)=O(|k|^{-a})$ for $|k|\to\infty$ \cite{Ochi1998}, which technically is not of compact support. Even so, waves with wavenumber $|k|\gg 1$ would carry very little energy -- and their physics would be predominantly surface tension and molecular effects, not hydrodynamics. So, from an ocean engineering point of view, applying a smooth cut-off to wavenumbers $|k|>K_M$ makes very little difference. Note furthermore that all the results would be uniform in $K_M.$
\end{remark}

}

\section{Strong solutions for the Alber equation} \label{sec:proofsexistence1}


To simplify notations we can rewrite equation  \eqref{eq:linearisedF_11} as
\beq \label{eq:wf1}
\begin{aligned} 
\partial_t f - 4\pi^2 i p k\cdot X f + \mathbb{B} [m,f]  = 0,  \qquad m(X,t)=\int_{\R^d} f(X,k,t)dk, \qquad f(X,k,0)=f_0(X,k), 
\end{aligned}
\ee
where 
\beq \label{eq:defBilBrack}
\begin{aligned} 
\mathbb{B} [m,f] =
i q \Big[P\Big(k-\frac{X}2\Big) - & P\Big(k+\frac{X}2\Big)\Big] m(X,t) + \\
&+ \epsilon i q \int_{\R^d}  m(y,t) \Big[f\Big(X-y,k-\frac{y}2,t\Big) - f\Big(X-y,k+\frac{y}2,t\Big)\Big] dy.
\end{aligned}
\ee

\begin{lemma}[Bounds on ${\mathbb{B}[m,f]}$]  \label{lm:aux1} Let $f,g,h\in L^1(\R^{2d}),$ $m\in L^1(\R^d)$ and consider ${\mathbb{B}[m,f]}$ as defined in equation \eqref{eq:defBilBrack}. Then
\beq\label{eq:aux1_lin200}
\|\mathbb{B}[m, f]\|_{L^1(\R^{2d})} \leqslant 2|q| \, \|P\|_{L^1(\R^d)} \|m\|_{L^1(\R^{d})}+2 |\epsilon q| \, \|m\|_{L^1(\R^{d})}\|f\|_{L^1(\R^{2d})}
\ee
and
\beq\label{eq:aux1_lin2}
\Big\|\mathbb{B}\Big[\int_{\R^d} f dk, f\Big]\Big\|_{L^1(\R^{2d})} \leqslant 2|q| \, \|P\|_{L^1(\R^d)} \|f\|_{L^1(\R^{2d})}+2 |\epsilon q| \, \|f\|_{L^1(\R^{2d})}^2.
\ee  
Moreover,
\beq\label{eq:aux1_LIps}
\begin{aligned}
& \Big \|\mathbb{B}\Big[\int_{R^d} g dk, g\Big]-\mathbb{B}\Big[\int_{\R^d} h dk, h\Big] \Big\|_{L^1(\R^{2d})}\leqslant \\ 
& \qquad \qquad \leqslant
2 |q| \, \Big(\|P\|_{L^1(\R^{d})}+ |\epsilon|\,\|g\|_{L^1(\R^{2d})}+|\epsilon|\,\|h\|_{L^1(\R^{2d})}\Big) \,\|g-h\|_{L^1(\R^{2d})}.
\end{aligned}
\ee
\end{lemma}

\noindent {\bf Proof:} For inequality \eqref{eq:aux1_lin200} observe that
\[
\begin{aligned}
\left\|\mathbb{B}[m,f]\right\|_{L^1(\R^{2d})} \leqslant & |q| \left\| 
\Big[P\Big(k-\frac{X}2\Big) -  P\Big(k+\frac{X}2\Big)\Big] m(X, y) \right\|_{L^1(\R^{2d})} +\\
& + |\epsilon q| \left\|\int_{\R^d}  m(y,t) \Big[f\Big(X-y,k-\frac{y}2,t\Big) - f\Big(X-y,k+\frac{y}2,t\Big)\Big] dy\right\|_{L^1(\R^{2d})}.
\end{aligned}
\]
We will treat each term separately. Firstly,
\[
\begin{aligned} 
\Big\| \Big[P\Big(k-\frac{X}2\Big) -  P\Big(k+\frac{X}2\Big)\Big] &  m(X,t) \Big\|_{L^1(\R^{2d})}  = \int_{\R^{2d}} \Big|P\Big(k-\frac{X}2\Big) - P\Big(k+\frac{X}2\Big)\Big| \, |m(X,y)|  dXdk  \\
& \leqslant \int_{\R^{2d}} \left|P\Big(k-\frac{X}2\Big)\right|  |m(X,t)| dXdk +
\int_{\R^{2d}} \left| P\Big(k+\frac{X}2\Big)\right|  |m(X)| dXdk \\
& =\int_{\R^{2d}} \left|P\Big(k-\frac{X}2\Big)\right|dk |m(X,t)| dX +
\int_{\R^{2d}} \left| P\Big(k+\frac{X}2\Big)\right|dk  |m(X,t)| dX\\
&=2 \|P\|_{L^1(\mathbb{R}^d)}  \|m\|_{L^1(\mathbb{R}^d)}. 
\end{aligned}
\]
Moreover
\[
\begin{aligned}
\Big\|\int_{\R^d}  m(y,t) \Big[f&\Big(X-y,k-\frac{y}2,t\Big) - f\Big(X-y,k+\frac{y}2,t\Big)  \Big] dy\Big\|_{L^1(\R^{2d})}    \\
&= \int_{\R^{3d}} 
|m(y,t)|  \left| f\Big(X-y,k-\frac{y}2,t\Big)  - f\Big(X-y,k+\frac{y}2,t\Big)\right| 
dy\, dX \, dk  \\
& \leqslant \int_{\R^{3d}}  \left| f\Big(X-y,k-\frac{y}2,t\Big) \right| dX \,dk \,| m(y,t)| \, dy +
\int_{\R^{3d}} \left| f\Big(X-y,k+\frac{y}2,t\Big) \right|  dX \,dk \, |m(y,t)| \, dy \\
& =2  \|f(t)\|_{L^1(\mathbb{R}^{2d})} \|m(t)\|_{L^1(\mathbb{R}^d)}.
\end{aligned}
\]
Inequality \eqref{eq:aux1_lin2} follows by virtue of  the elementary observation
\[
\|\int_{\R^d} f dk\|_{L^1(\R^d)} = \int_{\R^d} \left|\int_{\R^d} f(X,k,t) dk\right| dX \leqslant 
\int_{\R^{2d}} \left| f(X,k,t) \right| dk dX=\|f\|_{L^1(\R^{2d})}.
\]
For inequality \eqref{eq:aux1_LIps} we expand
\[
\begin{aligned}
\mathbb{B} & [\int_{\R^d} g dk,g]  - \mathbb{B} [\int_{\R^d} h dk,h] = 
i q \Big[P\Big(k-\frac{X}2\Big) - P\Big(k+\frac{X}2\Big)\Big] \Big( \int_{\R^d} g(X,k)dk -\int\limits_k h(X,k)dk \Big)  \\
&+ \epsilon i q \int_{\R^d} \int_{\R^d} g(y,k)dk \Big[g\Big(X-y,k-\frac{y}2\Big) - g\Big(X-y,k+\frac{y}2\Big)-h\Big(X-y,k-\frac{y}2\Big) + h\Big(X-y,k+\frac{y}2\Big)\Big] dy\\
&+ \epsilon i q \int_{\R^d} \Big( \int_{\R^d} g(y,k)dk - \int_{\R^d} h(y,k)dk \Big)\Big[h\Big(X-y,k-\frac{y}2\Big) - h\Big(X-y,k+\frac{y}2\Big)\Big] dy.
\end{aligned}
\]

The result follows by treating each term as before.
\qed

Moreover, consider some $s\in \mathbb{N}$ and the multi-indices $|\alpha+\beta+\gamma+\delta|\leqslant s.$ Let us denote
\[
f^{\alpha,\beta,\gamma,\delta} := X^\alpha k^\beta \partial_X^\gamma \partial_k^\delta f.
\]
By direct computation one obtains
\beq
\label{eq:aux123}
\begin{aligned} 
&\partial_t f^{\alpha,\beta,\gamma,\delta} - 4\pi^2 i p k\cdot X f^{\alpha,\beta,\gamma,\delta}   = \mathbb{B}^{(\alpha,\beta,\gamma,\delta)} [f],  \\ 
&f^{\alpha,\beta,\gamma,\delta}(X,k,0)=f^{\alpha,\beta,\gamma,\delta}_0(X,k):=X^\alpha k^\beta \partial_X^\gamma \partial_k^\delta f_0(X,k).
\end{aligned} 
\ee
The detailed expression for $\mathbb{B}^{(\alpha,\beta,\gamma,\delta)} [f]$ can be found in Appendix \ref{sec:appdeiff}, and it contains terms of the form $f^{\alpha',\beta',\gamma',\delta'},$ $X^{\alpha'}\partial_X^{\gamma'} m(X,t),$ for $\alpha'+\beta'+\gamma'+\delta'\leqslant  \alpha+\beta+\gamma+\delta.$ Furthermore, one can directly -- if somewhat tediously -- obtain the following 
\begin{lemma}[Bound on the nonlinearity $\mathbb{B}^{(\alpha,\beta,\gamma,\delta)}{[f]}$] \label{lm:bdabcdnonli} Let $1\leqslant s\in\N,$  and consider multi-indices
\[
|\alpha+\beta+\gamma+\delta| \leqslant s,
\]
and $\mathbb{B}^{(\alpha,\beta,\gamma,\delta)} [f]$ as in Appendix \ref{sec:appdeiff}. Assume also that $P \in \mathcal{S}(\R^d).$
Then there exists a $C=C(s,d,q,P)>0$ such that
\[
\|\mathbb{B}^{(\alpha,\beta,\gamma,\delta)}[f]\|_{L^1(\R^{2d})} \leqslant C \Big[1+\epsilon\|f\|_{\Sigma^{s-1,1}} \Big] \|f\|_{\Sigma^{s,1}}.
\]
\end{lemma}
\qed 

\subsection{Proof of Theorem~\ref{thrm:wfi_lwp}}\label{subsec:existyence}

Denote
\beq \label{eq:propww_1}
U(t): g(X,k) \mapsto e^{4\pi^2 i p k\cdot X t} g(X,k),
\ee
the free-space propagator, i.e. $g(t)=U(t)g_0$ means that
$
\partial_t g - 4\pi^2 i p k\cdot X g=0$ and $g(0)=g_0.$ 
Observe that, by construction, $\|U(t)g\|_{L^1_{X,k}}=\|g\|_{L^1_{X,k}}.$
Equation \eqref{eq:wf1} can now be written in mild form
\beq \label{eq:techmild1}
 f(X,k,t) = 
U(t)f_0 -  \int_{0}^t U(t-\tau) \mathbb{B}\Big[\int_{\R^d} f(\tau)dk, f(\tau)\Big] d\tau.
\ee
Define 
\[
\mathbb{E}:=\left\{ g \in L^\infty(0,T_0; L^1(\mathbb{R}^{2d})) \mbox{ so that } \|g\|_{L^\infty(0,T_0; L^1(\mathbb{R}^{2d}))} \leqslant M  \right\}
\]
for some $M,T_0>0$, to be determined below. Moreover denote 
\[
\mathbb{G}: \mathbb{E} \ni g \mapsto U(t)f_0 - \int_{0}^t U(t-\tau) \mathbb{B}\Big[\int_{\R^d} g(\tau)dk, g(\tau)\Big]d\tau.
\]

We will show that the operator $\mathbb{G}$ is a strict contraction on $\mathbb{E}.$
First we need to show  that  $\mathbb{G}\mathbb{E} \subseteq \mathbb{E}$.   Direct application of estimate  \eqref{eq:aux1_lin2} from Lemma \ref{lm:aux1} yields 
\[
\begin{aligned}
\|\mathbb{G} g\|_{L^\infty(0,T_0; L^1(\R^d))} & \leqslant \|f_0\|_{L^1} + T_0  \, \Big\| B\Big[\int_{\R^d} g(\tau)dk, g(\tau)\Big]\Big\|_{L^\infty L^1}\leqslant \\
& \leqslant \|f_0\|_{L^1} + T_0|q| \left( 2\|P\|_{L^1} \|g\|_{L^\infty L^1} + 2|\epsilon| \|g\|_{L^\infty L^1}^2 \right) \leqslant \|f_0\|_{L^1} + T_0|q|(2\|P\|_{L^1} M + 2|\epsilon| M^2). 
\end{aligned}
\]
A (non-sharp) way to guarantee that $\|\mathbb{G} g\|_{L^\infty_t L^1_{X,k}} \leqslant M$ is to consider
\beq \label{eq:T1}
M= 2\|f_0\|_{L^1(\R^{2d})} \quad \text{ and } \quad  T_0 < \frac{1}{|q|\, \max \{ 4\|P\|_{L^1(\R^d)}, \,\, 4|\epsilon|\} \,\, (M+1)}.
\ee
Now using  \eqref{eq:aux1_LIps} from Lemma \ref{lm:aux1}, for any $g,h\in\mathbb{E}$ we  obtain 
\[
\begin{aligned} 
\|\mathbb{G}g - \mathbb{G}h\|_{L^\infty L^1} & \leqslant T_0 \|B[\int\limits_k g(\tau)dk,g(\tau)]- B[\int\limits_k h(\tau)dk,h(\tau)]\|_{L^\infty_\tau L^1} \\
&\leqslant 2 T_0 |q| \, \Big(\|P\|_{L^1(\R^{d})}+ |\epsilon|\,\|g\|_{L^1(\R^{2d})}+|\epsilon|\,\|h\|_{L^1(\R^{2d})}\Big) \,\|g-h\|_{L^1(\R^{2d})}  \\
& \leqslant 2 T_0 |q| (\|P\|_{L^1(\R^d)}+2 \epsilon M) \|g-h\|_{L^1(\R^{2d})}.
\end{aligned} 
\]
For $T_0$ satisfying \eqref{eq:T1} the Lipschitz constant $L\leqslant T_0 |q| (\|P\|_{L^1}+ 2\epsilon M)$ of the mapping is strictly smaller than $1$.
Therefore, by virtue of the Banach Fixed Point Theorem, there exists a unique fixed point $f\in \mathbb{E},$ 
$f=\mathbb{G}f,$ i.e.\  a unique mild solution of \eqref{eq:wf1} for $t \in (0, T]$. Observe that by construction $\mathbb{G}g$ is continuous in time as a mapping with values in $L^1(\R^{2d})$.

Since $\|f(T_0)\|_{L^1(\R^{2d})}<\infty,$ we can repeat the argument and extend the solution in time. Thus the blowup alternative follows, i.e. either the solution exists for all times, or there exists a finite blow-up time $T<\infty$ so that $\lim\limits_{t\to T^-} \|f(t)\|_{L^1(\R^{2d})}=+\infty.$ Whether $T$ is finite or infinite, it will be called the {\em maximal time} for which $f(X,k,t)$ exists.


To show continuous dependence of solutions of \eqref{eq:wf1} on initial data we consider  $f(X,k,t)$ as above and   $g(X,k,t)$  being a solution of \eqref{eq:wf1} with initial data $g_0(X,k).$ Take some $T_1$ smaller than both the maximal times of $f$ and $g;$ then there exists some $M_1$
so that
$$\|f(t)\|_{L^\infty(0,T_1; L^1(\R^{2d}))}, \,\, \|g(t)\|_{L^\infty(0,T_1; L^1(\R^{2d}))} \leqslant M_1.$$
Now denote 
$h:=f-g;$
by subtracting the equations for $f,$ $g$ and using the same ideas as above, it follows that for all $t\in[0,T_1]$
\beq\label{eq:stability6t5r4e3w}
\begin{aligned}
\|h(t)\|_{L^1(\R^{2d})} \leqslant & \|f_0-g_0\|_{L^1(\R^{2d})} + 2\int_{0}^t \|h(\tau)\|_{L^1} \left( \|P\|_{L^1(\R^{d})}  + \epsilon (\|f(\tau)\|_{L^1(\R^{2d})} + \|g(\tau)\|_{L^1(\R^{2d})})  \right) d\tau \leqslant\\
\leqslant & \|f_0-g_0\|_{L^1(\R^{2d})} + 2(\|P\|_{L^1(\R^{d})}+2\epsilon M_1) \int_{0}^t \|h(\tau)\|_{L^1(\R^{2d})} d\tau. 
\end{aligned}
\ee
{Applying the the Gronwall inequality yields 
$$
\|h(t)\|_{L^1} \leqslant 
\|f_0-g_0\|_{L^1} \Big(1+ t 2(\|P\|_{L^1}+2\epsilon M_1) e^{t 2(\|P\|_{L^1}+2\epsilon M_1)}\Big) \qquad \forall t\in [0,T_1], 
$$
and hence the continuous dependence of solutions on initial data. 
}
\qed

\subsection{Propagation of regularity and Proof of Theorems~\ref{cor:RegSol} \& \ref{thrm:lwfi}} \label{subsec:propergef887}

\begin{theorem}[Local well-posedness  for the nonlinear Alber-Fourier-I equation on $\Sigma^{s,1}$] \label{thrm:NonlWellPosedReg} Denote    $f(X,k,t)$  the solution of \eqref{eq:wf1} with initial data $f_0(X,k)\in\Sigma^{s,1},$ \,\, $T=T(\|f_0\|_{L^1},q,\epsilon,\|P\|_{L^1})$  the maximal time for which $f(t)\in L^1(\R^{2d})$ and $M^0(t):=\|f(t)\|_{L^1(\R^{2d})}\in C[0,T).$ Moreover, for each $1\leqslant s\leqslant a_0$ denote
$
M^s(t):=\|f(t)\|_{\Sigma^{s,1}}.
$
Then there exist constant $C>0$ depending on $s,d,q,\epsilon,P$ and the background spectrum $P$ such that
\beq\label{eq:claim1jg}
M^s(t) \leqslant M^s(0) + C(s) \int_{0}^t  M^{s-1}(\tau)  M^s(\tau) d\tau \qquad \forall t\in[0,T),
\ee
and therefore, for all $s\in \N,$
\beq\label{eq:claim2jg}
M^s(t) < \infty \quad  \forall t \in [0,T), \qquad \quad f(t)\in C([0,T),\Sigma^{s,1}).
\ee
\end{theorem}

\noindent {\bf Proof:} Consider multi-indices $|\alpha+\beta+\gamma+\delta|\leqslant s;$ as was seen earlier, $f^{\alpha,\beta,\gamma,\delta} := X^\alpha k^\beta \partial_X^\gamma \partial_k^\delta f$ satisfies equation \eqref{eq:aux123}. By passing to mild form we have
\[
f^{\alpha,\beta,\gamma,\delta}(t)=U(t)f^{\alpha,\beta,\gamma,\delta}_0 + \int_{0}^t U(t-\tau)  \mathbb{B}^{(\alpha,\beta,\gamma,\delta)} [f(\tau)]  d\tau.
\]
Taking $L^1$ norms and using Lemmata~\ref{lm:aux1} and \ref{lm:bdabcdnonli} we have   
\begin{equation}\label{f_abgd}
 \|f^{\alpha,\beta,\gamma,\delta}(t)\|_{L^1} \leqslant  \|f^{\alpha,\beta,\gamma,\delta}_0\|_{L^1} + C \int_{0}^t  \|f(\tau)\|_{\Sigma^{s-1,1}} \|f(\tau)\|_{\Sigma^{s,1}}   d\tau.
\end{equation}
Equation \eqref{eq:claim1jg} follows by summing over all $|\alpha+\beta+\gamma+\delta|\leqslant s.$ 
The first part of equation \eqref{eq:claim2jg}  follows by applying recursively Gronwall's inequality to equation \eqref{eq:claim1jg}. The second part of equation \eqref{eq:claim2jg} follows automatically from the mild form \eqref{f_abgd} since the time integrals now are known to exist.
\qed

\bigskip

\noindent {\bf Proof of Theorem \ref{cor:RegSol}:} 
For the proof of regularity \eqref{eq:reg00122} it suffices to observe that the $\bigcap\limits_{s \in \mathbb{N}} \Sigma^{s,1}(\R^{2d})$ regularity is propagated in time by virtue of Theorem \ref{thrm:NonlWellPosedReg}, and that it implies Schwarz-class regularity by virtue of Lemma \ref{lm:Sigmasebeddings}.

For the proof of smoothness  with respect to the time variable stated in  \eqref{eq:reg00132z}, observe that upon applying the operator $\partial_t^l$ to equation \eqref{eq:wf1}, one obtains the problem
\[
\begin{aligned}
& \partial_t (\partial_t^l f) - 4\pi^2 i p k\cdot X (\partial_t^l f) + \mathbb{B}[m,\partial_t^l f]=\mathbb{B}_{(l)}[f], \qquad m(X,t)=\int_{\R^d} f(X,k,t)dk, \\ 
& \partial_t^l f(0) = 4\pi^2 i p k\cdot X (\partial_t^{l-1} f) - \mathbb{B}[m,\partial_t^{l-1} f]+\mathbb{B}_{(l-1)}[f],
\end{aligned}
\]
where
\[
\begin{aligned}
\mathbb{B}_{(l)}[f]&=-\epsilon i q \sum\limits_{0\leqslant l' < l} \binom{l}{l'}  
\int_{\R^d} \partial_t^{l-l'} m(y,t) \Big[\partial_t^{l'}f\Big(X-y,k-\frac{y}2,t\Big) - \partial_t^{l'}f\Big(X-y,k+\frac{y}2,t\Big)\Big] dy, \qquad 
\mathbb{B}_{(0)}[f]&=0.
\end{aligned}
\]
By working recursively in $l$ as in  the proof of Theorem \ref{thrm:NonlWellPosedReg}, the result follows.
\qed

\bigskip

\noindent {\bf Proof of Theorem~\ref{thrm:lwfi}:}  We start by recasting equation \eqref{eq:aux123} in mild form and taking the $L^1$ norm. 
Using the fact that $\epsilon=0$ and estimate in  Lemma \ref{lm:bdabcdnonli} we obtain 
\[
\|f^{\alpha,\beta,\gamma,\delta}(t)\|_{L^1(\R^{2d})} \leqslant \|f^{\alpha,\beta,\gamma,\delta}_0\|_{L^1(\R^{2d})} + C \int_{0}^t \|f(\tau)\|_{\Sigma^{s,1}(\R^{2d})}d\tau .
\]
Summing over all $|\alpha+\beta+\gamma+\delta|\leqslant s$  yields 
\[
\|f(t)\|_{\Sigma^{s,1}(\R^{2d})} \leqslant \|f_0\|_{\Sigma^{s,1}(\R^{2d})} + C \int_{0}^t \|f(\tau)\|_{\Sigma^{s,1}(\R^{2d})}d\tau.
\]
Then estimate  \eqref{eq:gfxclaim1yg} follows by Gronwall's inequality.

By virtue of Lemma \ref{lm:Sigmasebeddings}, for any $r$ and for $s'$ large enough we have $\|f(t)\|_{\Sigma^{r,\infty}
}\leqslant C \|f(t)\|_{\Sigma^{s',1}} \leqslant C e^{Ct}\|f_0\|_{\Sigma^{s',1}}.$

Now for the position density observe that
\[
\begin{aligned}
\|\widecheck{n}(t)\|_{L^\infty(\R^{d})}=
\sup\limits_{X\in\R^d}|\int_{\R^{d}} f(X,\xi,t)d\xi| \leqslant \int_{\R^{d}} \frac{d\xi}{\sup\limits_{X\in\R} (1+|\xi|^{d+1}) }\sup\limits_{X,\xi\in\R^d} |(1+|\xi|^{d+1}) f(X,\xi,t)| \\  \leqslant C \|f(t)\|_{\Sigma^{d+1,\infty}}\leqslant C' e^{C't}\|f_0\|_{\Sigma^{s',1}}.
\end{aligned}
\]
Moreover, considering equation \eqref{eq:LinAlbMOVETOSTART}  and using assumptions on $P$ we obtain 
\[
\|\partial_t f(t)\|_{L^\infty(\R^{2d})} \leqslant
 \| k\cdot X f(t)\|_{L^\infty(\R^{2d})} + C\|\widecheck{n}(t)\|_{L^\infty(\R^d)} \leqslant C\| f(t)\|_{\Sigma^{s_1,1}(\R^{2d})}, 
\]
for some $s_1\in\N$ large enough. Similarly
\[
\partial_t \widecheck{n}= \int_{R^d}\partial_t f(X,k,t)dk  = 4\pi^2 i p \int_{\R^d} k\cdot X f dk - i q \int_{\R^d}\Big[P\Big(k-\frac{X}2\Big) -  P\Big(k+\frac{X}2\Big)\Big]dk \int_{\R^d} f(X,\xi,t)d\xi 
\]
 implies
\[
\|\partial_t \widecheck{n}(t)\|_{L^\infty(\R^d)} \leqslant
C \| |k|^{d+2} |X| f(t)\|_{L^\infty(\R^{2d})} + C\|\widecheck{n}(t)\|_{L^\infty(\R^d)} \leqslant C\| f(t)\|_{\Sigma^{s_1',1}(\R^{2d})}, 
\]
for some $s_1'\in\N$ large enough.
Thus estimate \eqref{eq:gfxclaim2yg} follows by selecting $s_2=\max\{s',s_1,s_1'\}$.
\qed

\section{The free-space position density}

In this Section we will establish some properties of the free-space position density  {$n_f(x,t)$,    defined in  \eqref{eq:def_n_f},} that we will use for the proof of Theorem \ref{thrm:MainTheorem}. 

\begin{lemma}[Alternative expression for $\widecheck{n}_f.$]\label{lm:nfhat}

\beq
\widecheck{n}_f(X,t) := \mathcal{F}^{-1}_{x\to X}[n_f(x,t)] = \widecheck{w}_{0}(X,2\pi p t X), 
\ee
where  $\widecheck{w}_{0}(A,B)=\mathcal{F}^{-1}_{(x,k)\to (A,B)}[w_{0}(x,k)].$
\end{lemma}

\noindent {\bf Proof:} Simple calculations yield 
\[
\begin{aligned}
\widecheck{n}_f(X,t) &= \int_{\R} e^{2\pi i x\, X} n_f(x,t)dx = \int_{\R^2} e^{2\pi i x\, X} w_{0}(x-2\pi p kt,k) dkdx \\
&=\int_{\R^4} e^{2\pi i x\, X-2\pi i [A\, (x-2\pi p kt)+B\,  k]} \widecheck{w}_{0}(A,B) dkdx dA dB \\
&=\int_{\R^2}\int_{\R^2} e^{2\pi i x\,[X-A]} e^{2\pi i k\, [2\pi p t A-B]} dxdk \,\, \widecheck{w}_{0}(A,B) dA dB = \widecheck{w}_{0}(X,2\pi p t X).
\end{aligned}
\]
\qed

\begin{lemma}[Uniform bound for $X\widetilde{n}_f.$] \label{lem:unif_bound_force_laplaced}
Assume that there exists some $D>0$ such that
\[
|\widecheck{w}_{0}(X,K)| \leqslant \frac{D}{1+|X|^2+|K|^2}
\]
and  $\widetilde{n}_f(X,\omega)$ as in equation \eqref{eq:defntilde}.
Then, there exists a constant $C>0$ such that for all $X\in \R$ 
\[
\sup\limits_{\real \omega>0} |X\widetilde{n}_f(X,\omega)| \leqslant C\,  D.
\]
\end{lemma}

\noindent {\bf Proof:} 
Using Lemma \ref{lm:nfhat} one readily checks that
\[
\begin{aligned}
\sup\limits_{\real \omega>0} |X\widetilde{n}_f(X,\omega)|
& = \sup\limits_{\real \omega>0} \int_{0}^\infty |e^{-\omega t} X\widecheck{n}_f(X,t)|dt \leqslant \int_{0}^\infty |X\widecheck{n}_f(X,t)|dt\\
&=\int_{0}^\infty |X\widecheck{w}_{0}(X,2\pi p t X)|dt \leqslant 
\int_{0}^\infty \Big|X\frac{D}{1+|X|^2+|2\pi p t X|^2}\Big|dt\leqslant \int_{0}^\infty \frac{CD}{1+|2\pi p t|^2}dt.
\end{aligned}
\]

\begin{observation} \label{obs:f0_w0_regularity} We will use assumptions of the form
\[
|\widecheck{w}_{0}(X,K)| \leqslant \frac{D_r}{1+|X|^r+|K|^r}
\]
in the sequel, which are weaker versions of $\widecheck{w}_{0}\in\Sigma^{r,\infty}(\R^{2d})$. 
By virtue of Lemma \ref{lm:Sigmasebeddings} it follows that, for some $r'$ large enough
\[
D_r \leqslant \|\widecheck{w}_0\|_{\Sigma^{r,\infty}(\R^{2d})} \leqslant C\|{w}_0\|_{\Sigma^{r',\infty}(\R^{2d})}.
\]
\end{observation}

\begin{lemma}[Space-time $L^2$ estimates for the free-space position density]\label{lm:freespace1} Let 
\[
|\widecheck{w}_{0}(X,K)| \leqslant \frac{D_r}{1+|X|^r+|K|^r} 
\]
for some large enough $r$ and constant $D_r>0$. Assume moreover
$r-\frac{1}2>a>b\geqslant0$
($a,b,r$ don't have to be integer.) Then
\[
\left( \int\limits_{X,t} |X^at^b \widecheck{n}_f(X,t)|^2 dXdt \right)^{\frac{1}2} \leqslant C(a,b) {D_r}.
\]
\end{lemma}

\noindent {\bf Proof:} We will break up the norm as follows:
\begin{figure}[h!]
\begin{center}
\includegraphics[width=0.75\textwidth]{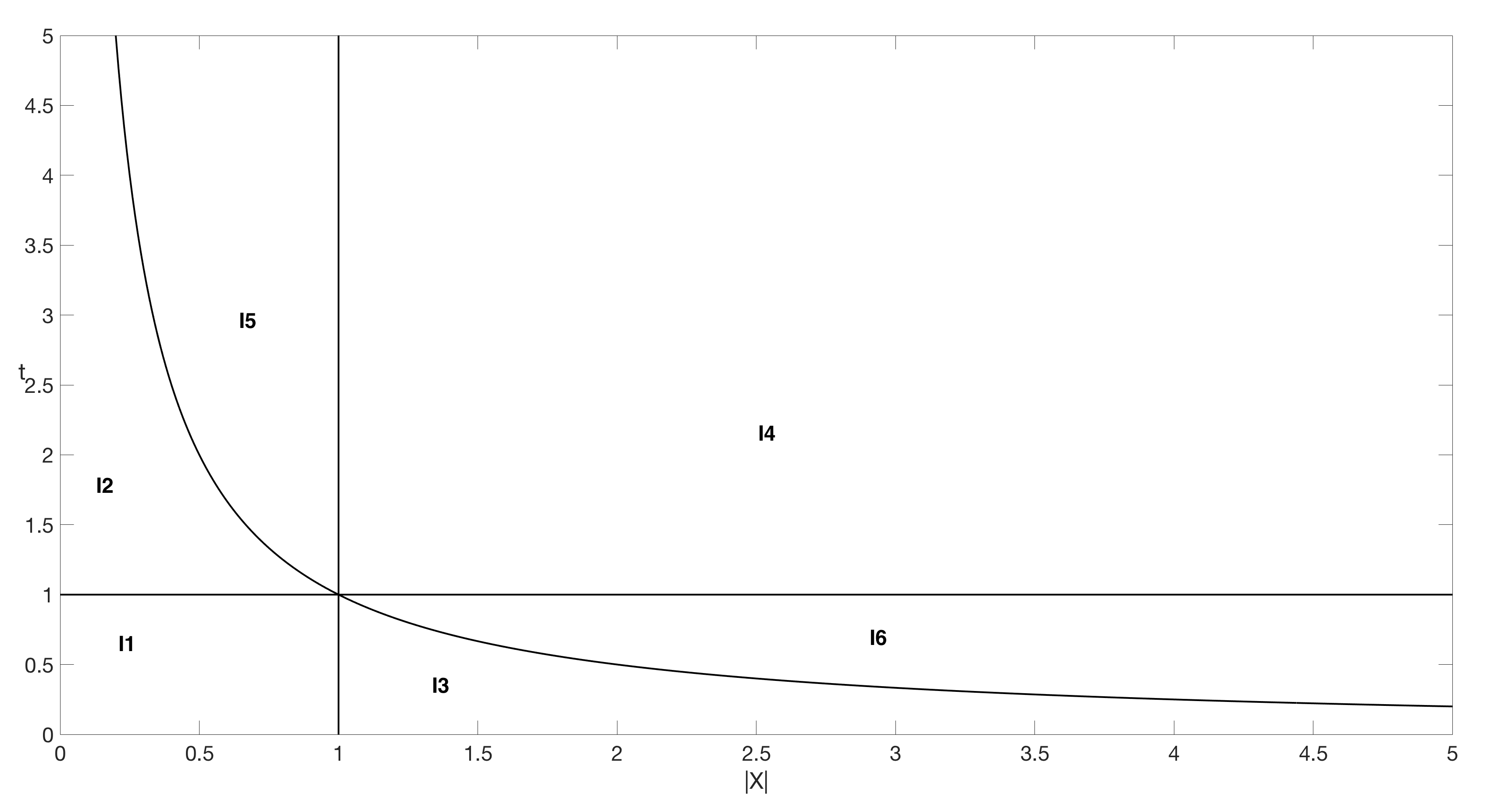}	
\caption{The domains of integration for the integrals $I_j,$ $j=1,\dots,6.$}
\label{Fig:Is}
\end{center}
\end{figure}
\[
\bac
\|X^at^b \widecheck{n}_f(X,t)\|_{L^2_{X,t}} = \int\limits_{t,X} |X^at^b\widecheck{w}_{0}(X,2\pi p t X)|^2 dXdt = I_1+I_2+I_3+I_4+I_5+I_6=\\
=\int\limits_{|X|<1,0<t<1} + 
\int\limits_{|X|<1,1<t<1/|X|} +
\int\limits_{|X|>1,0<t<1/|X|} +
\int\limits_{|X|>1,1<t} +
\int\limits_{1/t<|X|<1,1<t} +
\int\limits_{|X|>1,1/|X|<t<1},
\ea
\]
see Figure \ref{Fig:Is}.
One readily observes that
\[
\begin{aligned}
I_1&=\int_{\substack{|X|<1}} \int_{|t|<1} |X^at^b\widecheck{w}_{0}(X,2\pi p t X)|^2 dtdX \leqslant C D_r^2,
\\
I_2&= \int_{|X|<1} \int_{1}^{1/|X|}  |X^at^b\widecheck{w}_{0}(X,2\pi p t X)|^2 dtdX  \leqslant 
D_r^2 C\int_{|X|<1} |X|^{2a}\int_{1}^{1/|X|} t^{2b}dt dX\\
& \hspace{ 6cm }  \leqslant
 D_r^2 C\int_{|X|<1} |X|^{2a-2b-1}  dX \leqslant D_r^2C, 
\\
I_3&= \int_{|X|>1} \int_{0}^{1/|X|}  |X^at^b\widecheck{w}_{0}(X,2\pi p t X)|^2 dtdX \leqslant
D_r^2 \int_{|X|>1} \int_{0}^1 \frac{|X|^{2a}}{|X|^{2r}} dtdX=  CD_r^2 \int_{1}^\infty X^{2(a-r)} dX \leqslant C D_r^2.  
\end{aligned} 
\]
By using the elementary  observation that $\frac{x^{2a}t^{2b}}{(1+x^r+(xt)^r)^2} \leqslant \frac{x^{2a}t^{2b}}{(xt)^{2r} } = x^{2(a-r)}t^{2(b-r)}$  for $t\geq 0$, $x \geq 0$, we have 
\[
I_{4} = \int_{1}^\infty \int_{|X|\geq 1}  |X^at^b\widecheck{w}_{0}(X,2\pi p t X)|^2 dtdX\leqslant CD_r^2\int_{1}^\infty \int_{|X|\geq 1}  {|X|^{2(a-r)}t^{2(b-r)}} dtdX \leqslant CD_r^2 
\]
and 
\[
\begin{aligned}
I_{5}
& = \int_{1}^\infty \int_{|X|=1/t}^1  |X^at^b\widecheck{w}_{0}(X,2\pi p t X)|^2 dXdt \leqslant CD_r^2\int_{1}^\infty \int_{|X|=1/t}^1  |X|^{2(a-r)} t^{2(b-r)} dXdt
\\
& = CD_r^2\int_{1}^\infty t^{2(b-r)}  \int_{|X|=1/t}^1  |X|^{2(a-r)}  dXdt \leqslant 
C'D_r^2\int_{1}^\infty t^{2(b-r)} (1-t^{-2(a-r)-1}) dt
\\
& \leqslant  C'D_r^2 \Big(\int_{1}^\infty t^{2(b-a)-1} dt + \int_{1}^\infty t^{2(b-r)} dt\Big)  \leqslant CD_r^2.
\end{aligned}
\]
Finally, by using the elementary  observation that $\frac{x^{2a}t^{2b}}{(1+x^r+(xt)^r)^2} \leqslant \frac{x^{2a}t^{2b}}{x^{2r} } = x^{2(a-r)}t^{2b}$ we have 
\[
\begin{aligned}
I_{6} &=
\int_{1}^\infty \int_{1/|X|}^1  |X^at^b\widecheck{w}_{0}(X,2\pi p t X)|^2 dtdX \leqslant 
CD_r^2 \int_{1}^\infty \int_{0}^1 t^{2b} |X|^{2(a-r)}  dtdX\\
& = CD_r^2 \int_{1}^\infty  |X|^{2(a-r)} dX \int_{0}^1 t^{2b}  dt \leqslant C'D_r^2.
\end{aligned}
\]
{Collecting all above estimates yield the result stated in the lemma. }
\qed

We will see that, in the stable case, the position density for the linearised problem inherits these estimates in an appropriate sense.

\section{Proof of Theorem~\ref{thrm:MainTheorem}}\label{sec:proof}

\subsection{The Laplace transform picture}\label{sec:5.1}

Theorem  \ref{thrm:lwfi}  implies that the Laplace transforms $\widetilde{n}(X,\omega),$  is well-defined and analytic for $\real(\omega)$ large enough. Moreover we {can apply} Fubini to the effect that $\widetilde{n} = \mathcal{L}[\int\limits_{\R^d} f dk]= \int\limits_{\R^d}\mathcal{L}[f] dk,$ for $\real(\omega)$ large enough. The same follows for $\widetilde{n}_f(X,\omega)$ by setting $P(k)=0.$

Thus if we first take the Laplace transform of equation \eqref{eq:LinAlbMOVETOSTART}, 
\begin{eqnarray}
\nonumber
\omega \widetilde{f} - f_0(X,k) - 4\pi^2 i p k \cdot X \widetilde{f} 
+qi \Big[ P\Big(k-\frac{X}2\Big) - P\Big(k+\frac{X}2\Big) \Big] \widetilde{n}(X,\omega)=0,
\end{eqnarray} 
re-arrange terms 
\[
f(X,k,\omega) = \frac{f_0(X,k) - qi \left[ P(k-\frac{X}2) - P(k+\frac{X}2) \right] \widetilde{n}(X,\omega)}{\omega - 4\pi^2 i p k \cdot X},
\]
and integrate in $k$  we obtain
\beq\label{eq:0linearisedF_113}
\widetilde{n}(X,\omega) =
\int\limits_{\R^d}\frac{f_0(X,k)}{\omega - 4\pi^2 i p k \cdot X} dk -qi
\int\limits_{\R^d} \frac{P(k-\frac{X}2) - P(k+\frac{X}2)}{\omega - 4\pi^2 i p k \cdot X}dk \,\,\cdot \,\,\widetilde{n}(X,\omega).
\ee
This is exactly the first expression in equation \eqref{eq:laplastartpoint}.  From this alternative derivation  we obtain that for $X\neq 0$ and  $d=1$
\beq 
\widetilde{n}_f(X,\omega)=\int_\R\frac{f_0(X,k)}{\omega - 4\pi^2 i p k \cdot X} dk=\frac{1}{4\pi i pX}\mathbb{H}[f_0(X,\cdot)]\Big(\frac{\omega}{4\pi^2 i pX}\Big)
\ee 
and
\beq
\label{eq:h_htilde}
\widetilde{h}(X,\omega)=  qi
\int_\R \frac{P(k+\frac{X}2) - P(k-\frac{X}2)}{\omega - 4\pi^2 i p k \, X}dk=\frac{q}{4\pi p} \mathbb{H}[D_XP(\cdot)]\Big(\frac{\omega}{4\pi^2 i p X}\Big).
\ee

\begin{observation}[Case $X=0$]\label{obs:Case_X_is_0}
For $X=0$ we have $\widetilde{h}(0,\omega)=0,$ and $\widetilde{n}_f(0,\omega)=\frac{1}\omega \int_\R f_0(0,k)dk,$ which is of course consistent with Lemma \ref{lm:nfhat} and its consequence $\widecheck{n}_f(0,t)=\widecheck{w}_0(0,0).$ Thus it follows that $\widecheck{n}(0,t)=\widecheck{n}_f(0,t)=\widecheck{w}_0(0,0)$ for all $t.$ 
\end{observation}

\begin{observation}[Domain of analyticity \& Sokhotski-Plemelj]\label{obs:Domain_of_Analyticity} From the above explicit expressions it follows that, for each $X\in\R,$ the Laplace transforms $\widetilde{h}(X,\omega),$ $\widetilde{n}(X,\omega),$ $\widetilde{n}_f(X,\omega)$ are analytic in $\omega$ for all $\real(\omega)>0.$

Moreover,  for $X\neq 0,$ we have
\beq\label{eq:sokh_plemelj_appl1}
\widetilde{H}(X,s):=\lim\limits_{\eta\to 0} \widetilde{h}(X,\eta+is) = \lim\limits_{\eta\to 0}\frac{q}{4\pi p} \mathbb{H}[D_X P]\Big(\frac{\eta+is}{4\pi^2 i p X}\Big) =\frac{q}{4\pi p}  \mathbb{S}[D_X P]\Big(\frac{s}{4\pi^2 p X}\Big)
\ee
and
\beq\label{eq:sokh_plemelj_appl2}
\bac
\widetilde{N}_f(X,s):=\lim\limits_{\eta\to 0} \widetilde{n}_f(X,\eta+is) = \lim\limits_{\eta\to 0}\frac{1}{4\pi i pX} \mathbb{H}[f_0(X,\cdot)](\frac{\eta+is}{4\pi^2 i p X}) 
=\frac{1}{4\pi i pX} \mathbb{S}[f_0(X,\cdot)](\frac{s}{4\pi^2  p X})
\ea
\ee
by virtue of the Sokhotski-Plemelj formula, cf. Theorem \ref{thrm:SokhPlem} in the Appendix.
Moreover, observe that
\beq\label{eq:obs_Hpenrose_hPenrose}
|1-\widetilde{h}(X,\omega)|\geqslant \kappa \quad \forall X\in\R, \,\real(\omega)>0 \qquad \Rightarrow \qquad |1-\widetilde{H}(X,s)|\geqslant \kappa \quad  \forall X,s\in \R.
\ee
\end{observation}

\subsection{Inverting the Laplace Transform}

Recalling now equation \eqref{eq:laplastartpoint}, we set
\[
\widetilde{I}(X,\omega):=\frac{\widetilde{h}(X,\omega)}{1-\widetilde{h}(X,\omega)} X\widetilde{n}_f(X,\omega);
\]
then
\[
X\widecheck{n}(X, t) - X\widecheck{n}_f(X, t) = \mathcal{L}^{-1}_{\omega\to t}[\widetilde{I}(X,\omega)].
\] 
Observe also that equation \eqref{eq:sokh_plemelj_appl1} and \eqref{eq:sokh_plemelj_appl2} imply
\[
I(X,s):=\lim\limits_{\eta\to 0^+} \widetilde{I}(X,\eta+is) = \frac{ \widetilde{H}(X,s)}{1-\widetilde{H}(X,s)}X\widetilde{N}_f(X,s).
\]
We will  use Theorem \ref{thrm:InvLaplaceOPen} from the Appendix to compute $\mathcal{L}^{-1}_{\omega\to t}[\widetilde{I}(X,\omega)]$ for each $0\neq X\in\R$. To that end, we will need to check that its assumptions are satisfied, namely  that $\widetilde{I}(X,\omega)$ is bounded and analytic on $\{\real(\omega>0\});$ that $|\widetilde{I}(X,\omega)|$ decays uniformly as $|\omega|\to\infty;$ and that $I(X,\cdot)\in L^1(\R) \cap C^0(\R)$ for all $X\in\R;$

First of all, Observation \ref{obs:Domain_of_Analyticity} directly implies that $\widetilde{I}(X,\omega)$ is bounded and analytic on the open half-plane $\{\real(\omega>0\}).$ Moreover,
\[
\lim\limits_{\rho\to\infty}\sup\limits_{\substack{\real(\omega)>0\\ |\omega|>\rho}} |\widetilde{I}(X,\omega)| \leqslant 
\sup\limits_{\real(\omega')>0} |X\widetilde{n}_f(X,\omega')| \,\,\cdot \,\,
\lim\limits_{\rho\to\infty}\sup\limits_{\substack{\real(\omega)>0\\ |\omega|>\rho}} |\widetilde{h}(X,\omega)|= 0
\]
where in the last step we used Lemma \ref{lm:decay_htilde} from the Appendix. 

Finally, {the expression for $I(X,s)$ implies that it is continuous in $s.$ To show that  ${I}(X,\cdot) \in L^1(\R)$ uniformly in $X$} observe that
\[
\int_{\R} |I(X,s)|ds \leqslant \frac{1}\kappa \, \sup\limits_{s'\in\R} |X\widetilde{N}_f(X,s')| \, \int_{\R} |\widetilde{H}(X,s)|ds \leqslant C, 
\]
where we used property \eqref{eq:obs_Hpenrose_hPenrose}, Lemma \ref{lem:unif_bound_force_laplaced} for $X\widetilde{N}_f,$ and Theorem \ref{thrm:l1hilbertzerointegral} for $\widetilde{H}$ (observe in particular that, by construction, $D_XP$ is a function of compact support with integral $\int_{\R} D_XP(k)dk=0$ for all $0\neq X\in\R,$ hence Theorem \ref{thrm:l1hilbertzerointegral} indeed applies).

So all the assumptions of Theorem \ref{thrm:InvLaplaceOPen} are satisfied, and we can apply it to the effect that
\beq\label{eq:workasin}
X\widecheck{n}(X, t) - X\widecheck{n}_f(X, t) =  
 \int_{-\infty}^\infty   e^{i\, s\, t} \frac{\widetilde{H}(X,s)}{1-\widetilde{H}(X,s)} X\widetilde{N}_f(X,s) ds.
\ee

\begin{remark}
If one tries to use Theorem \ref{thrm:InvLaplaceOPen} directly on $X\widetilde{n}(X,\omega)$ then the only way to guarantee the {$L^1$-property required in} equation \eqref{eq:L1constraint} seems to be requiring $\int_{k\in\R} f_0(X,k) dk=0$ for all $X\in\R.$ Here instead we only require that {the limit as  $\eta\to 0^+$ of  the difference $X\widetilde{n}(X,s+i\eta) - X\widetilde{n}_f(X,s+i\eta)$ is in  $L^1_s(\R)$} uniformly in $X$, avoiding any extra assumptions on the initial data.
\end{remark}

\subsection{Space-time estimates for the force}

First we use equation \eqref{eq:workasin} to prove the following estimate

\begin{lemma}\label{lm:ntonfforsomeindices}
Let $a>1,$ $b>0,$
and moreover recall that, since $f_0\in\mathcal{S}(\R^2),$
\[
|\widecheck{w}_{0}(X,K)| \leqslant \frac{ \|w_0\|_{\Sigma^{r',\infty}} }{1+|X|^r+|K|^r} 
\]
for any $r,$  in particular for $r>\max\{ a+\frac{1}2, b+\frac{1}2\}.$
Then there exists a $C=C(a,b,P)$ so that
\[
\| t |X|^{a} \widecheck{n}\|_{L^2_{X,t}} + \||X|^b \widecheck{n}\|_{L^2_{X,t}} \leqslant \frac{ C(a,b,P) D_r}{\kappa}
\]
for all $0\neq X\in\R.$
Note that, by virtue of Observation \ref{obs:f0_w0_regularity}, $D_r\leqslant C \|w_0\|_{\Sigma^{r',\infty}}$ for some $r'$ sufficiently large.
\end{lemma}

\noindent {\bf Proof:} First we will bound $X^at^b \widecheck{n}$ norms from appropriate quantities involving $\widecheck{n}_f$. Using the alternate Fourier transform $\mathfrak{F}$, introduced in Section \ref{subsec:defs_notations},  from \eqref{eq:workasin} follows
\[
X\widecheck{n}(X,t) - X\widecheck{n}_f(X,t) =  \mathfrak{F}^{-1}_{s\to t} \left[ \frac{X\widetilde{N}_f(X,s)  \widetilde{H}(X,s)}{1-\widetilde{H}(X,s)} \right] 
\]
This implies 
\[
\begin{aligned}
\|X^b\widecheck{n} - X^b\widecheck{n}_f \|_{L^2_{X,t}}=\Big\|
\mathfrak{F}^{-1}_{s\to t} \left[ \frac{X^b\widetilde{N}_f(X,s)  \widetilde{H}(X,s)}{1-\widetilde{H}(X,s)} \right] \Big\|_{L^2_{X,t}}=
 C \Big\| \frac{X^b\widetilde{N}_f(X,s) \widetilde{H}(X,s)}{1-\widetilde{H}(X,s) } \Big \|_{L^2_{X,s}} \\
\leqslant C \Big(\sup\limits_{X,s} \Big|\frac{1}{1-\widetilde{H}(X,s)}\Big|\Big) \, \Big(\sup\limits_{X,s} |\widetilde{H}(X,s)|\Big) \, 
\| X^b\widetilde{N}_f \|_{L^2_{X,s}} 
\end{aligned} 
\]
For the first factor we use equation \eqref{eq:obs_Hpenrose_hPenrose}. For the second factor observe that, by virtue of Theorem \ref{thrm:HilbReg}, we have 
\beq\label{eq:extraaa}
\sup\limits_{X,s} |\widetilde{H}(X,s)|=|\frac{q}{4\pi p}| \sup\limits_{\zeta,t} |\mathbb{S}[D_\zeta P](t)| \leqslant C \sup\limits_\zeta\|\mathbb{S}[D_\zeta P]\|_{H^1} \leqslant C' \sup\limits_\zeta \|D_\zeta P\|_{H^1} \leqslant C'',
\ee
so finally 
\beq\label{eq:spacetime_n_1}
\|X^b\widecheck{n}\|_{L^2_{X,t}} \leqslant C \| X^b\widetilde{N}_f \|_{L^2_{X,s}} =
C' \| X^b\widecheck{n}_f \|_{L^2_{X,t}}
\ee
since $\widecheck{n}_f=\mathfrak{F}^{-1}[\widetilde{N}_f].$
{Now working similarly and using \eqref{eq:workasin} we have}
\[
X^at\big[\widecheck{n}(X,t) - \widecheck{n}_f(X,t)\big] =  i \int_{-\infty}^\infty   e^{ist} \partial_s  \frac{X^a\widetilde{N}_f(X,s) \widetilde{H}(X,s)}{1-\widetilde{H}(X,s)} ds, 
\]
which implies 
\beq\label{eq:newnumb223}
\begin{aligned}
& \|X^at(\widecheck{n}(X,t) - \widecheck{n}_f(X,t))\|_{L^2_{X,t}}=C\Big \|\partial_s  \frac{X^a\widetilde{N}_f(X,s) \widetilde{H}(X,s)}{1-\widetilde{H}(X,s)}\Big\|_{L^2_{X,s}} 
\\
& \leqslant C \|\partial_s X^a\widetilde{N}_f\|_{L^2_{X,s}} \sup\limits_{X,s} \Big|\frac{\widetilde{H}(X,s)}{1-\widetilde{H}(X,s)}\Big| 
+ C \left(\int_{\R^2} |X^a\widetilde{N}_f(X,s)|^2 \Big[\sup\limits_{s'} \Big|\frac{\partial_s \widetilde{H}(X,s')}{(1-\widetilde{H}(X,s'))^2}\Big|\Big]^2 ds dX\right)^{\frac{1}2}.
\end{aligned} 
\ee
{ 
Now, observe that
\[
\|\partial_s X^a\widetilde{N}_f\|_{L^2_{X,s}} = \|t X^a\widecheck{n}_f\|_{L^2_{X,t}}
\]
by virtue of a Fourier transform;  $|\widetilde{H}(X,s)/(1-\widetilde{H}(X,s))| \leqslant C'''/\kappa^2$ by virtue of \eqref{eq:obs_Hpenrose_hPenrose} and  \eqref{eq:extraaa};
and
\beq\label{eq:the1overXcomesin}
\begin{array}{c}
\Big|\frac{\partial_s \widetilde{H}(X,s)}{(1-\widetilde{H}(X,s))^2}\Big| \leqslant \frac{C}{\kappa^2} \Big|\partial_s \widetilde{H}(X,s)\Big| 
=\frac{C'}{\kappa^2} \Big|\partial_s   \mathbb{S}[D_X P]\Big(\frac{s}{4\pi^2 p X}\Big)\Big| =
\frac{C''}{\kappa^2} \frac{1}{|X|}\Big|\mathbb{S}[D_X P']\Big(\frac{s}{4\pi^2 p X}\Big)\Big|
\end{array}
\ee
so that by collecting all this and inserting it back in \eqref{eq:newnumb223} we get
\[
\|X^at(\widecheck{n}(X,t) - \widecheck{n}_f(X,t))\|_{L^2_{X,t}}
\leqslant \frac{C}{\kappa}\|t X^a\widecheck{n}_f\|_{L^2_{X,t}} + \frac{C}{\kappa^2} \sup\limits_{s'} \Big| \mathbb{S}[D_XP'](\frac{s'}{4\pi^2 p X}) \Big| \|X^{a-1}\widecheck{n}_f\|_{L^2_{X,t}}.
\]
{Using our assumptions on $P$ we have}
\[
\sup\limits_{s' \in \R} | \mathbb{S}[D_XP']\Big(\frac{s'}{4\pi^2 p X}\Big)| \leqslant 
\sup\limits_{\zeta, \tau} | \mathbb{S}[D_\zeta P'](\tau)|\leqslant \sup\limits_{\zeta\in \R}\| D_\zeta P'\|_{H^1(\R)} \leqslant C,
\]
and therefore 
\beq
\label{eq:spacetime_n_2}
\|tX^a\widecheck{n}\|_{L^2_{X,t}} \leqslant C\left(\frac{1}{\kappa}+\frac{1}{\kappa^2}\right) \left( \|t X^a\widecheck{n}_f\|_{L^2_{X,t}} + \|X^{a-1}\widecheck{n}_f\|_{L^2_{X,t}} \right)
\ee
}
{Then the  result of the lemma follows by combining estimates \eqref{eq:spacetime_n_1} and \eqref{eq:spacetime_n_2}  together with  Lemma~\ref{lm:freespace1}.}
\qed 

{Applying now Lemma~\ref{lm:ntonfforsomeindices}   we obtain  estimate~\eqref{eq:landaydampingstatement}  stated in Theorem~\ref{thrm:MainTheorem}.}

\subsection{Construction of  the wave operator}\label{subsec:proofs_main}

Equation \eqref{eq:LinAlbMOVETOSTARTmild} implies
\beq \label{eq:jhgfc1112}
\begin{aligned}
e^{-4\pi^2 i p k \cdot X t}  f(X,k,t) -  f_0(X,k)= &J(X,k,t),\\
J(X,k,t):=& qi \int_{0}^t e^{-4\pi^2 i p k\cdot X \tau}  
\frac{P\Big(k+\frac{X}2\Big) - P\Big(k-\frac{X}2\Big)}{X} X \widecheck{n}(X,\tau)d\tau
\end{aligned} 
\ee
For any  $0<\theta<1/2$ and  $\gamma>1$ using  the Cauchy-Schwarz inequality we have 
\[
\begin{aligned}
& \int_{\R} |J(k,X,t)| dX \leqslant
C \sup\limits_{\zeta,s} |D_{\zeta}P(s)| \int\limits_{\R} \int_{0}^t |X\widecheck{n}(X,\tau)| d\tau dX
 \\
& \leqslant C' \int_{0}^{+\infty} \int_{0}^t \frac{\sqrt{1+(|X|^\theta \tau)^{2} + |X|^{2\gamma} }}{\sqrt{1+(|X|^\theta \tau)^{2} + |X|^{2\gamma} }} |X\widecheck{n}(X,\tau)| d\tau dX \\
& \leqslant C'' \sqrt{ \int_{\R} \int_{0}^t \Big[1+(|X|^\theta \tau)^{2} + |X|^{2\gamma} \Big] |X|^2 |\widecheck{n}(X,\tau)|^2 d\tau dX} \cdot
\sqrt{ \int_{0}^{+\infty} \int_{0}^t \frac{1}{1+(|X|^\theta \tau)^{2} + |X|^{2\gamma} } d\tau dX}.
\end{aligned} 
\]
The first factor  in the last estimates  is estimated by
\[
\begin{aligned}
& \sqrt{ \int_{0}^{+\infty} \int_{0}^t \Big[1+(|X|^\theta \tau)^{2} + |X|^{2\gamma} \Big] |X\widecheck{n}(X,\tau)|^2 d\tau dX} \leqslant \|
(1+|X|^\theta t + |X|^\gamma)X \widecheck{n}
\|_{L^2_{X,t}} \\
& \leqslant \| X\widecheck{n}\|_{L^2_{X,t}} +  \| t|X|^{1+\theta}\widecheck{n}\|_{L^2_{X,t}} +  \| |X|^{1+\gamma}\widecheck{n}\|_{L^2_{X,t}} \leqslant C(\theta,\gamma,P)\|w_0\|_{\Sigma^{r',\infty}}
\end{aligned}
\]
for some $r'$ large enough by virtue of Lemma \ref{lm:ntonfforsomeindices}.
For the other factor we break the integral up over the contributions from different regions,
\[
\int_{0}^{+\infty} \int_{0}^t \frac{1}{1+(|X|^\theta \tau)^{2} + |X|^{2\gamma} } d\tau dX = I_1+I_2+I_3+I_4+I_5+I_6,
\]
where we use the same breakdown as in Figure~\ref{Fig:Is}. Without loss of generality we assume $t>1.$ Then the first integral is estimates as
\[
I_1\leqslant \int_{0}^1 \int_{0}^1  d\tau dX = 1.
\]
{For the second integral we have}
\[
I_2\leqslant \int_{1}^\infty \int_{0}^{1/t} \frac{1}{x^{2\theta}\tau^2} d\tau dx =
  \int_{1}^\infty \tau^{-2} \int\limits_{x=0}^{1/t} x^{-2\theta} d\tau dx=
C  \int_{1}^\infty \tau^{-2+2\theta-1} d\tau = C (1+t^{-2+2\theta})  \leqslant C'.
\]
{Here we used $-2\theta >-1 \iff \theta<1/2$ for the integral with respect to $x$ to exist  and $-3+2\theta<-1 \iff \theta <1$ for the integral  with respect to $\tau$ to exist.}
Moreover
\[
I_3\leqslant  \int_{1}^\infty \int_{0}^{1/x} x^{-2\gamma} dt dx =   \int_{1}^\infty  x^{-2\gamma-1} dx =C,
\]
where we used $-2\gamma-1<-1 \iff \gamma>0$.
For $I_4$ we refer to Lemma \ref{lm:myyoyngineq} in the Appendix, where setting $\zeta=3/4$ leads to $$\frac{1}{(x^\theta \tau)^2 + x^{2\gamma}} \leqslant \frac{1}{(x^\theta \tau)^{\frac{3}2} x^{\frac{\gamma}2}}=\tau^{-\frac{3}2} x^{-\frac{\gamma}2-\frac{3}{2}\theta}.$$ Thus
\[
I_4 \leqslant C \int_{1}^\infty \int_{1}^t \tau^{-\frac{3}2} x^{-\frac{\gamma}2-\frac{3}{2}\theta} d\tau dx = C  \,\, \left( \tau^{-\frac{1}2}\Big|_{1}^{t} \right)\,\, \left(x^{1-\frac{\gamma}2-\frac{3}{2}\theta}\Big|_{1}^\infty\right) = C'(1+t^{-\frac{1}2}), 
\]
where we used the fact that, by assumption, $\gamma/2+3\theta/2>5/4>1.$
{The next integral is estimated as}  
\[
\begin{aligned}
I_5 \leqslant C \int_{0}^1 \int_{1/x}^\infty \frac{1}{x^{2\theta}t^2} dtdx = C \int_{0}^1 x^{-2\theta} \int_{1/x}^\infty \tau^{-2}d\tau dx 
\\
=C \int_{0}^1 x^{-2\theta}\,\, \left( \tau^{-1}\Big|_{1/x}^\infty \right) dx 
= C \int_{0}^1 x^{1-2\theta} dx =C', 
\end{aligned} 
\]
since $1-2\theta>-1 \iff \theta<1/2.$
Finally, 
\[
I_6 \leqslant C \int_{1}^\infty \int_{1/x}^1 x^{-2\gamma} d\tau dx \leqslant \int_{1}^\infty x^{-2\gamma-1} dx \leqslant C.
\]
So we showed that
\[
\int_{\R} |J(k,X,t)| dX \leqslant C \|w_0\|_{\Sigma^{r',\infty}}.
\]
Since $J(X,k,t)$ is an {  absolutely convergent} integral in $t,$ the uniform-in-$t$ bound automatically implies the existence of
\[
J^\infty(X,k):=
\lim\limits_{t\to\infty} J(X,k,t)=qi \int_{0}^\infty e^{-4\pi^2 i p k\cdot X \tau}  
\frac{P\Big(k+\frac{X}2\Big) - P\Big(k-\frac{X}2\Big)}{X} X \widecheck{n}(X,\tau)d\tau \in L^\infty(\R; L^1(\R)).
\]
Now equation \eqref{eq:jhgfc1112} can be recast as
\[
 U(-t)f(t) - f_0= J(t) \qquad \Rightarrow \qquad \lim\limits_{t\to\infty} \big(  U(-t)f(t) - f_0 \big) = J^\infty.
\]
By setting $\mathbb{W}(w_0) := \mathcal{F}_{X\to x} [f_0 + J^\infty],$ we have
\[
\| w(t)-E(t)\mathbb{W}(w_0)\|_{L^\infty(\R^2)} \leqslant 
\|f(t)-U(t)\big(f_0+J^\infty\big)\|_{L^\infty(\R;  L^1(\R))}=
\|U(-t)f(t)-f_0-J^\infty \|_{L^\infty(\R; L^1(\R))},
\]
hence equation \eqref{eq:waveoperatorstatement} follows.

{ 
\begin{remark}
Observe that by collecting the above it follows that
\[
\| \mathbb{W}(w_0)\|_{L^\infty_{x,k}} \leqslant  \|w_0\|_{L^\infty_{x,k}} + \|J^\infty\|_{L^\infty_k L^1_X} \leqslant C' \|w_0\|_{\Sigma^{r',\infty}}.
\]	
\end{remark}
}

\section{Proof of Theorem \ref{thrm:PenroseConditionResolve}}\label{sec:PenrCond765}

{In this section we present the proof of the last main results.  We split the proof into  four parts. }

\subsection{Elaboration and symmetry of (A).}
Assuming condition (A) holds,  there exists a sequence
  $(X_n,\omega_n)=(X_n,a_n+ib_n)\in\R\times\{\real(z)>0\}$  such that $\lim\limits_{n\to \infty} \widetilde{h}(X_n,\omega_n)=1.$
Without loss of generality we can assume $X_n\neq 0$ for all $n\in\N$ (it suffices to observe that $\widetilde{h}(0,\omega)=0$ for all $\omega$). Note that   $X_*$ can still be zero.

\noindent{\bf Symmetry:} {The expression for $\tilde h(X, \omega)$ in  \eqref{eq:h_htilde}, i.e.\  $\widetilde{h}(X,\omega)=  qi
\int_\R \frac{P(k+\frac{X}2) - P(k-\frac{X}2)}{\omega - 4\pi^2 i p k \, X}dk,$    yields  that, for $X_n,a_n,b_n\in\R$ as above, we have the following equivalence 
\[
\lim\limits_{n\to \infty}\widetilde{h}(X_n,a_n+ib_n)= 1 \iff \lim\limits_{n\to \infty}\widetilde{h}(X_n,-a_n+ib_n)=1, 
\]
i.e.
\[
\exists X_n\in\R,\, \omega_n\in\C \,\,:\,\, \widetilde{h}(X_n,\omega_n)\to 1 \quad \iff \quad 
\exists X_n\in\R,\, \real(\omega_n)\geqslant 0 \,\,:\,\, \widetilde{h}(X_n,\omega_n)\to 1.
\]
Indeed all the  conditions $(A), (B)$ and $(C)$ have this symmetry.
}

\noindent{\bf Claim I:} The sequence $(X_n,\omega_n)$ is bounded.

\noindent{\bf Proof:} If $|X_n|+|\omega_n|\to \infty,$ then 
$$\lim\limits_{n\to \infty}\widetilde{h}(X_n,\omega_n)=qi\lim\limits_{n\to \infty}
\int_\R \frac{P(k+\frac{X_n}2) - P(k-\frac{X_n}2)}{\omega_n - 4\pi^2 i p k \, X_n}dk=
0\neq 1.
$$
Thus $(X_n,\omega_n)$ has accumulation points in $\R\times\{\real(z)\geqslant 0\}$ and  from now on we will denote
\beq
(X_*,a_*+ib_*)=(X_*,\omega_*):=\lim\limits_{n\to\infty}(X_n,\omega_n),
\ee
up to extraction of a subsequence.

\noindent{\bf Claim II:} 
Denote
\beq
\Omega_n:=\frac{\omega_n}{4\pi p i X_n}=\frac{b_n - i a_n}{4\pi p X_n}.
\ee 
Then $\Omega_n$ is bounded.

\noindent{\bf Proof of the claim:} First of all observe that $\Omega_n$ is well-defined since, as we saw above, $X_n\neq 0.$
By virtue of equation \eqref{eq:h_htilde}, 
\beq
\widetilde{h}(X_n,\omega_n)=\frac{q}{4\pi p}\mathbb{H}[D_{X_n}P](\Omega_n). 
\ee
Clearly, if $|\Omega_n|\to\infty$ then $(q/4\pi p)\mathbb{H}[D_{X_n}P](\Omega_n)\to 0\neq 1.$
Thus, by extracting yet another subsequence if necessary, we have $(X_n,\Omega_n) \to (X_*,\Omega_*) \in \R \times\C.$

\subsection{Proof of $\boldsymbol{(A) \iff (B).}$}

\noindent{\bf Case 1:} If $\imag(\Omega_*)\neq 0$ then, by continuity,
\[
\widetilde{h}(X_n,\omega_n)\to 1 \iff \frac{q}{4\pi p} \mathbb{H}[D_{X_*}P](\Omega_*)=1.
\] 
\noindent{\bf Case 2:} If $\imag(\Omega_*)=0$ then, by the Sokhotski-Plemelj formula (cf. Theorem \ref{thrm:SokhPlem}), for $X_*> 0$ we have
\[
\widetilde{h}(X_n,\omega_n)\to 1 \iff \frac{q}{4\pi p} \mathbb{S}[D_{X_*}P](\Omega_*)=1 \iff
\left\{
\bac
\frac{q}{4\pi p} \mathbb{H}[D_{X_*}P](\Omega_*)=1, \\
i \frac{q}{4\pi p}D_{X_*}P(\Omega_*)=0, 
\ea
\right\} 
\]
while for $X_*<0$ we have $\overline{\mathbb{S}[D_{X_*}P](\Omega_*)}=1,$ leading to the same end result. For $X_*=0$ observe that both one-sided limits $\imag(\Omega_n)\to 0^\pm,$ yield the same result as well.

\medskip
Checking that (B) implies (A) is obvious.

\subsection{Proof of $\boldsymbol{(B) \iff (C).}$}

Denote $\mathbb{F}_{X}(\Omega):=\mathbb{H}[D_{X}P](\Omega).$ Like before, if $\pm X_*>0$ we have $\pm\imag(\Omega_*)<0,$ and for $X_*=0$ we should take each one-sided limit separately. All these cases follow the same steps, so without loss of generality we only present the case $X_*>0.$

Assume Case 1 of (B) above holds, i.e. $\exists X_*>0, \imag(\Omega_*)\neq 0$ such that $\mathbb{H}[D_{X_*}P](\Omega_*)={4\pi p}/{q}$. 

Then by virtue of the argument principle \cite{Penrose1960}, for any contour $\gamma$ within the lower half-plane containing $\Omega_*,$ its image  $\mathbb{F}_{X_*}(\gamma):=\{ z| \exists w\in \gamma : z=\mathbb{F}_X(w)\}$ is enclosing $4\pi p/q.$ Let us  select  $\gamma_\eta$ the closed contour comprised by parts of  the horizontal line $\R-i \eta$ and the semicircle $\{ \frac{e^{i\theta}}{\eta}, \; \; \theta\in(-{\pi},0)\}$.  Clearly, $\Omega_*$  will eventually be enclosed by $\gamma_\eta$ for $\eta$ small enough, thus $\mathbb{F}_{X_*}(\gamma_\eta)$ is enclosing $4\pi p/q$ for $\eta$ small enough. Using the  decay properties of $\mathbb{F}_{X_*}(\omega)$ as $|\omega|\to\infty$ (cf. Lemma \ref{lm:decay_htilde} in the Appendix) and the Sokhotski-Plemelj formula, it follows that $\lim\limits_{\eta\to 0} \mathbb{F}_{X_*}(\gamma_\eta)=\Gamma_X$ as defined in equation \eqref{eq:defcurveG}, i.e. $4\pi p/q \in \overset{\circ}\Gamma_{X_*}$.

If Case 2 of (B) above holds, denote $\Omega_n$ a sequence of points on $\Gamma_{\eta_n}$ such that $\Omega_n\to \Omega_*;$ then by construction $\lim\limits_{n\to\infty} \mathbb{F}_{X_*}(\Omega_n)=4\pi p/q$ and therefore $4\pi p/q \in \lim\limits_{\eta\to 0}\mathbb{F}_{X_*}(\gamma_\eta)=\Gamma_{X_*}.$

To prove that $(C) \implies (B),$ first we need to observe that, since $\lim\limits_{|X|\to\infty}\|D_XP\|_{H^1}=0,$ there exists $M>0$ such that for $|X|>M$ all points of $\Gamma_X$ are inside $\{ z\in \C|\, |z|<2\pi p/q\}.$ 
Thus $4\pi p/q \in \overline\Gamma$ implies $\exists X_*\in[-M,M]$ such that $d(4\pi p/q,\overset{\circ}\Gamma_{X_*}).$ One now readily checks that there exists $\Omega_*$ with $\imag(\Omega_*)\leqslant 0$ such that $\lim\limits_{\substack{\Omega\to\Omega_*\\\imag(\Omega)<0}}\mathbb{F}_{X_*}(\Omega)=4\pi p/q.$

\subsection{Sufficient condition for stability}

This follows from the elementary observation that, for the curve $\Gamma_X$ on the complex plane, which starts and ends at $0,$ to be winding around the real number $4\pi p/q,$ it is necessary to intersect the real axis somewhere on the right of $4\pi p/ q.$ The argument can be {easily  adapted for limiting case $4\pi p/q \in \Gamma_X$.} See also Figures \ref{Fig:1}, \ref{Fig:3} for a visualisation of this point.

The proof is completed by observing that, according to equation \eqref{eq:defcurveG}, $\Gamma_X=\{\mathbb{S}[D_XP](t), t\in \R\}$ intersects the real axis only for those $t_*$ that are quasi-critical points, $D_XP(t_*)=0.$

\section{Applications}\label{sec:conclusions}

{ 

\subsection{The question: Are realistic sea states modulationally (un)stable?}

Landau damping for the Alber equation (i.e. dispersion of inhomogeneities in the presence of a homogeneous background) has been conjectured at least since \cite{Onorato2003}, but  no precise results existed before {the one presented here}. In this paper we establish rigorously the decay of inhomogeneities in the stable case, but for ocean engineers the most immediate question is a practical and reliable way to investigate whether a given spectrum is stable or not.

 Alber's ``eigenvalue relation''  is a system of two (real valued) nonlinear equations in three (real) unknowns, which in general has one-dimensional manifolds of solutions.  Determining whether such a system has solutions or not is not straightforward, and has attracted a lot of attention in the ocean waves community \cite{Gramstad2017,Onorato2003,Ribal2013a,Stiassnie2008}.  In  \cite{Gramstad2017} a state-of-the-art investigation of this question is presented, describing the challenges. 
We will show that criterion (C) of Theorem \ref{thrm:PenroseConditionResolve} provides a reliable and more straightforward way to investigate the modulational stability of any given spectrum. But first let us go over how we choose the spectra to be investigated.

\subsection{JONSWAP spectra and the North Atlantic Scatter Diagram}

While the power spectrum of a sea state can in principle be directly measured, in practice often  parametric spectra are used. A widely used such parametric spectrum is the so-called JONSWAP spectrum (the initials stand for ``Joint North Sea Wave Project'', and some typical profiles can be found in Figure \ref{Fig:exjons}),
\beq\label{eq:defeqJONSWpsp}
S_{\alpha,\gamma,k_0}(k)=S(k) = \frac{\alpha}{2k^3}e^{-\frac{5}4 (\frac{k_0}{k})^2} \gamma^{exp[ -(1-\sqrt{k/k_0})^2 / 2\delta^2 ]}, \qquad \delta=\delta(k)=\left\{
\begin{array}{ll}
0.07, & k\leqslant k_0,\\
0.09, & k> k_0.
\end{array}
\right.
\ee
This was introduced in \cite{hasselmann1973measurements} following extensive study of measured nonparametric spectra, and it incorporates several physical insights: it is effectively zero in a neighbourhood of $k=0,$ it has a power-law decay for $k\gg 1,$ and it is unimodal.  The free parameters  are  $\alpha>0$, which increases with the power of the sea state (i.e. larger $\alpha$ leads to larger significant wave height $H_s$),  $\gamma>1$ which increases with the ``peakiness'' of the spectrum (i.e. larger $\gamma$ leads to more peaked spectra, with larger $H_s$ as well), and $k_0$ stands for the peak wavenumber. Very often a JONSWAP spectrum is fitted to a time-series of point measurements for the frequency $\omega$ (instead of the wavenumber $k$) but, assuming unidirectional propagation, the conversion between a wavenumber-resolved and frequency-resolved spectra is standard \cite{Ochi1998}.
\begin{figure}[h!]
\begin{center}
\includegraphics[width=0.59\textwidth]{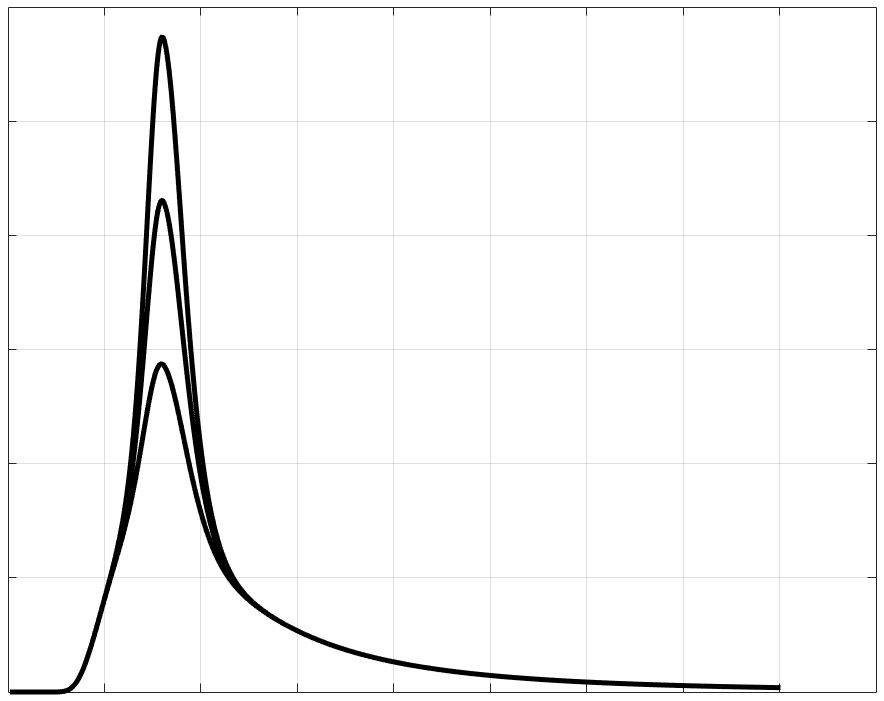}
\caption{Some common profiles of JONSWAP spectra.
}
\label{Fig:exjons} 
\end{center}
\end{figure}
It is widely used in the study of realistic sea states, e.g. \cite{Gramstad2017,Ribal2013a,ThemistoklisSapsis2016,DetNorskeVeritasGermanischerLloyd2017}, as it offers an intuitive and plausible parametrization of spectra in terms of power, peakiness and carrier wavenumber. 

Now the question becomes, what are some realistic values for $\alpha,\gamma$ and $k_0$ corresponding to various plausible scenarios in the ocean? A canonical data set has been created precisely in this context; it is called the North Atlantic Scatter Diagram \cite[ p. 244]{DetNorskeVeritasGermanischerLloyd2017}, and it includes measured statistics from 100000 sea states in the North Atlantic, along with the likelihood for each sea state. A JONSWAP spectrum (i.e. $\alpha,\gamma$ and $k_0$ values) can then be fitted to each sea state  using state of the art engineering practice \cite[ Section 3.5.5]{DetNorskeVeritasGermanischerLloyd2017}. The fact that parameter values are fitted and not measured directly has a few implications: for example, several sea states end up having $\gamma=1$ (smallest allowed value) or $\gamma=5$ (largest allowed value). More importantly, it is a priori possible that we could end up with some modulationally unstable spectra through this route. In contrast, if power spectra were measured directly, it doesn't seem likley that an ``unstable spectrum'' could be robustly measured at all.

So now it should be clear how we choose the spectra to investigate: we will work with JONSWAP spectra, fitted to the North Atlantic Scatter Diagram according to the state of the art \cite{DetNorskeVeritasGermanischerLloyd2017}. Ultimately each blue star in Figure \ref{Fig:synoptic} corresponds to one such JONSWAP spectrum,  and it has a known likelihood of being observed at a random point in the North Atlantic, at a random time of the year (this likelihood is not plotted here, but can be found in \cite{DetNorskeVeritasGermanischerLloyd2017}).

\subsection{Implementation}\label{sec:8.3}

One should start with the important observation that, for the question of modulation instability of JONSWAP spectra, $k_0$ happens to play no role\footnote{We would like to thank A. Babanin and O. Gramstad for their helpful insights on this point.}. This is well-known \cite{Gramstad2017,Ribal2013a}, but we will  demonstrate  it for completeness.

Let us begin with
 Alber's eigenvalue relation for some JONSWAP spectrum $S(k)$, e.g. as  in equation (2) of \cite{Gramstad2017}:
\beq\label{gralnhjfctrd}
\exists \Omega\in\C, \,\, X\in\R  \qquad 
1 + \omega_0k_0^2 \int\limits_{k\in\R} \cfrac{ S(k+\frac{X}2)-S(k-\frac{X}2) }{\Omega+ \frac{\omega_0}{4k_0^2 }kX} dk=0
\ee 
Recall that the existence of such $\Omega,X$ means the spectrum is unstable. 
By rescaling
$
X'= {X}/{k_0},$
$\Omega' := -\Omega{4k_0}/(X\omega_0),$ and changing variable $k'=k/k_0$
 problem \eqref{gralnhjfctrd} is seen to be equivalent to 
\beq\label{eq:OdinALbCond}
\exists \Omega' \in \C,\,\, X'\in\R  \qquad  \mathbb{H} [D_{X'} P](\Omega') =\frac{1}{4\pi}
\ee
where 
\beq
P(k) = \frac{\alpha}{2k^3}e^{-\frac{5}4 k^{-2}} \gamma^{exp[ -(1-\sqrt{k})^2 / 2\delta^2 ]}, \qquad \delta=\delta(k)=\left\{
\begin{array}{ll}
0.07, & k\leqslant 1,\\
0.09, & k> 1,
\end{array}
\right.
\ee
is the JONSWAP spectrum with $k_0=1$ and the original $\alpha,\gamma.$ So the original value of $k_0$ will play no further role in checking stability.

To actually do the checking, 
we recall part (C) of Theorem \ref{thrm:PenroseConditionResolve}: instability exists if and only if
\beq\label{eq:scaledcontourcr}
\frac{1}{4\pi} \mbox{ is on, or enclosed by, the curve }  \Gamma_X:=\lim\limits_{\eta\to0^+} \mathbb{H}[D_XP](t-i\eta) \mbox{ for some }  X\in \R.
\ee
So instead of checking for the existence of solutions of a system of nonlinear equations, we simply check whether $1/4\pi$ is on, or enclosed by, a curve in the complex plane. The computation of the curve itself is somewhat demanding, since it involves a very nearly singular integral. Still, it can be done much more reliably and quickly than   checking for existence of solutions of \eqref{gralnhjfctrd}.

After some numerical testing, it is found sufficient to  approximate
\[
\Gamma_X(t)=\lim\limits_{\eta\to0}\mathbb{H}[D_X P(\cdot)] (t-i \eta) \approx 
\mathbb{H}[D_X P(\cdot)] ({t -i \texttt{tol}}), \qquad \texttt{tol=1e-4}.
\]
In all relevant cases here we observe that condition \eqref{eq:scaledcontourcr} is satisfied if and only if it is satisfied for $X=0$ (and this seems to be the case for any unimodal spectrum). 
Once we generate an approximation to $\Gamma_X,$ the built-in MATLAB function \texttt{inpolygon} is then used to determine if the target $4\pi p/q$ is contained in $\Gamma_X \cup\{0\}.$

Application
 to individual spectra is  vizualized in Figures \ref{Fig:1} and \ref{Fig:3}. 
Synoptic plots showing the stable and unstable regions of the $\gamma-\alpha$ plane can be found in Figure \ref{Fig:synoptic}.
 There is broad agreement with \cite{Gramstad2017,Ribal2013a}, but we find somewhat fewer unstable sea states. Modulationally unstable sea states are the prime suspects for rogue waves \cite{Bitner-Gregersen2015,Athanassoulis2017,ThemistoklisSapsis2016,Dematteis2017,Gramstad2018,Onorato2013}, and we find that such sea states are very unlikely but nevertheless they do exist, with an estimated total likelihood of $\approx 2\cdot 10^{-3}$. This is broadly consistent with the record of observations of rogue waves.

\subsection{The bifurcation from Landau damping to modulation instability}

Another aspect of practical interest is to 
understand the bifurcation
 from  stability  to instability e.g. as  $\alpha$ or $\gamma$ increases. This has been  thought of as a violent change in behavior once a borderline stable spectrum became unstable.   Such a change in behavior is the object of  numerical experiments in \cite{Janssen2003}, where it is noted that instead only a gradual transition is found. In fact, the lack of a dramatic bifurcation was seen as a challenge for the validity Alber equation in the aformentioned works. However our proof here (and the heuristic results of \cite{Athanassoulis2017} for the unstable case) show that  indeed the Alber equation  only predicts  a gradual transition. 
 
For example, assume $\gamma_*,\alpha_*$ are exactly on the separatrix of the stable/unstanle regions as in Figure \ref{Fig:synoptic}. Also take $(\gamma_m,\alpha_m)$ a sequence of points in the stable region with $\lim\limits_{m\to \infty} (\gamma_m,\alpha_m) = (\gamma_*,\alpha_*).$ Now denote $S_*(k)=S_{\alpha_*,\gamma_*,k_0}(k),$
$S_m(k)=S_{\alpha_m,\gamma_m,k_0}(k).$ For each $S_m(k)$ we have Landau damping, and dispersion of inhomogeneities over a timescale controlled by $\kappa_m.$ However, as $m\to\infty$  it takes longer and longer for the inhomogeneities  to disperse; this can be seen e.g. by considering equation \eqref{eq:laplastartpoint}, which in this case becomes
\[
X\widetilde{n}_m(X,\omega) - X\widetilde{n}_f(X,\omega) = \frac{\widetilde{h}_m(X,\omega)}{1-\widetilde{h}_m(X,\omega)} X\widetilde{n}_f(X,\omega),
\]
assuming the same initial inhomogeneity for all $m.$
So when $m \to \infty$ we have $\kappa_m \to 0$ and the force decays more and more slowly, until it ceases to have any time decay at all. 

On the other hand, in the unstable case a very slow rate of growth would make the instability irrelevant; moreover, a very small bandwidth of unstable wavenumbers $X$ would make the resulting extreme events supported over unrealistically large regions (e.g. thousands of wavelengths) \cite{Athanassoulis2017}; but there are no energy transport mechanisms to support such events. In other  words, to really observe the modulation instability a fast enough rate of growth and a large enough bandwidth of unstable wavenumbers are required. 

So a barely stable and a barely unstable spectrum would lead to very similar behaviour over physically relevant timescales and lengthscales, reconciling the findings of \cite{Janssen2003} with the analysis of the Alber equation.

\begin{figure}[ht!]
\begin{center}
\includegraphics[width=0.33\textwidth]{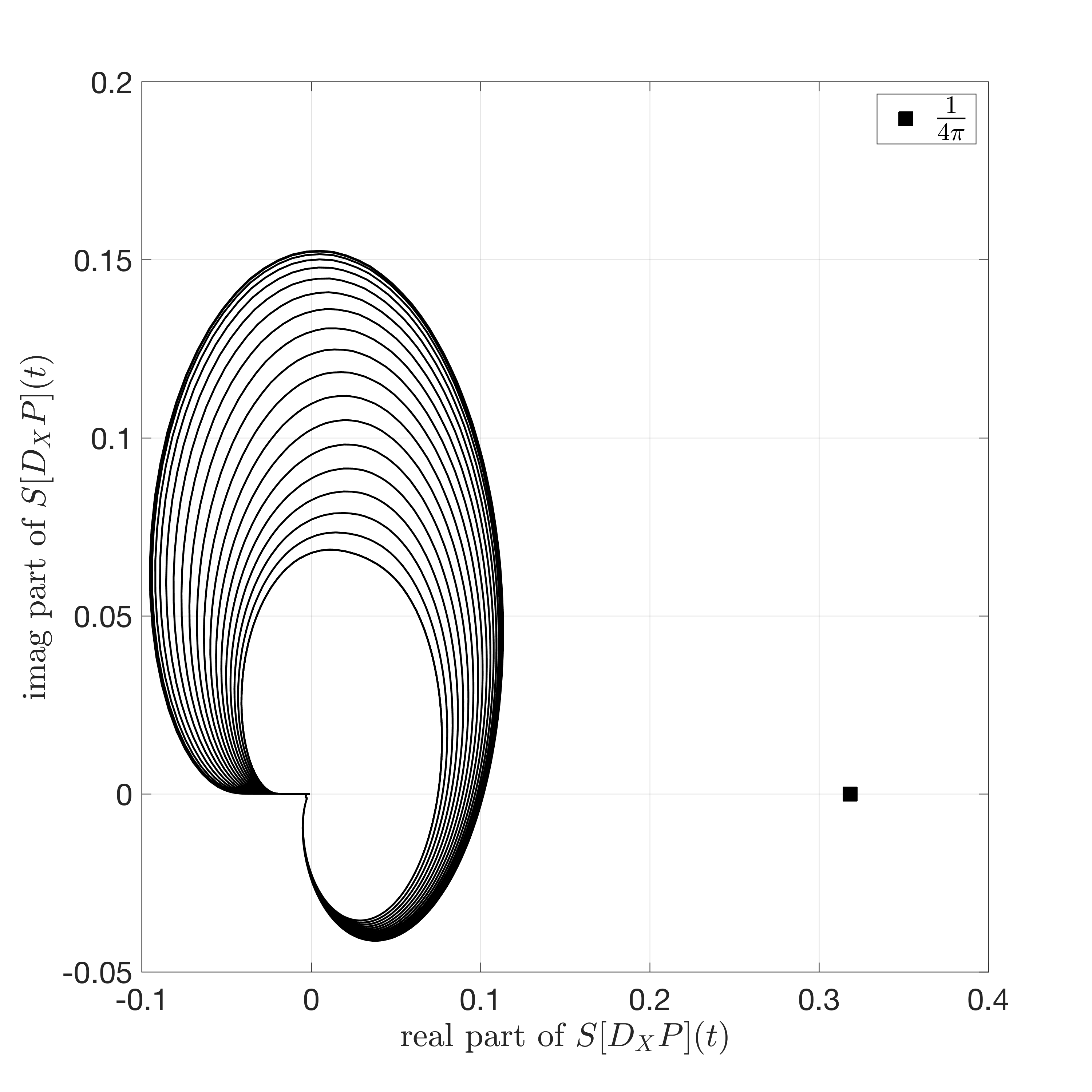} 
\includegraphics[width=0.66\textwidth]{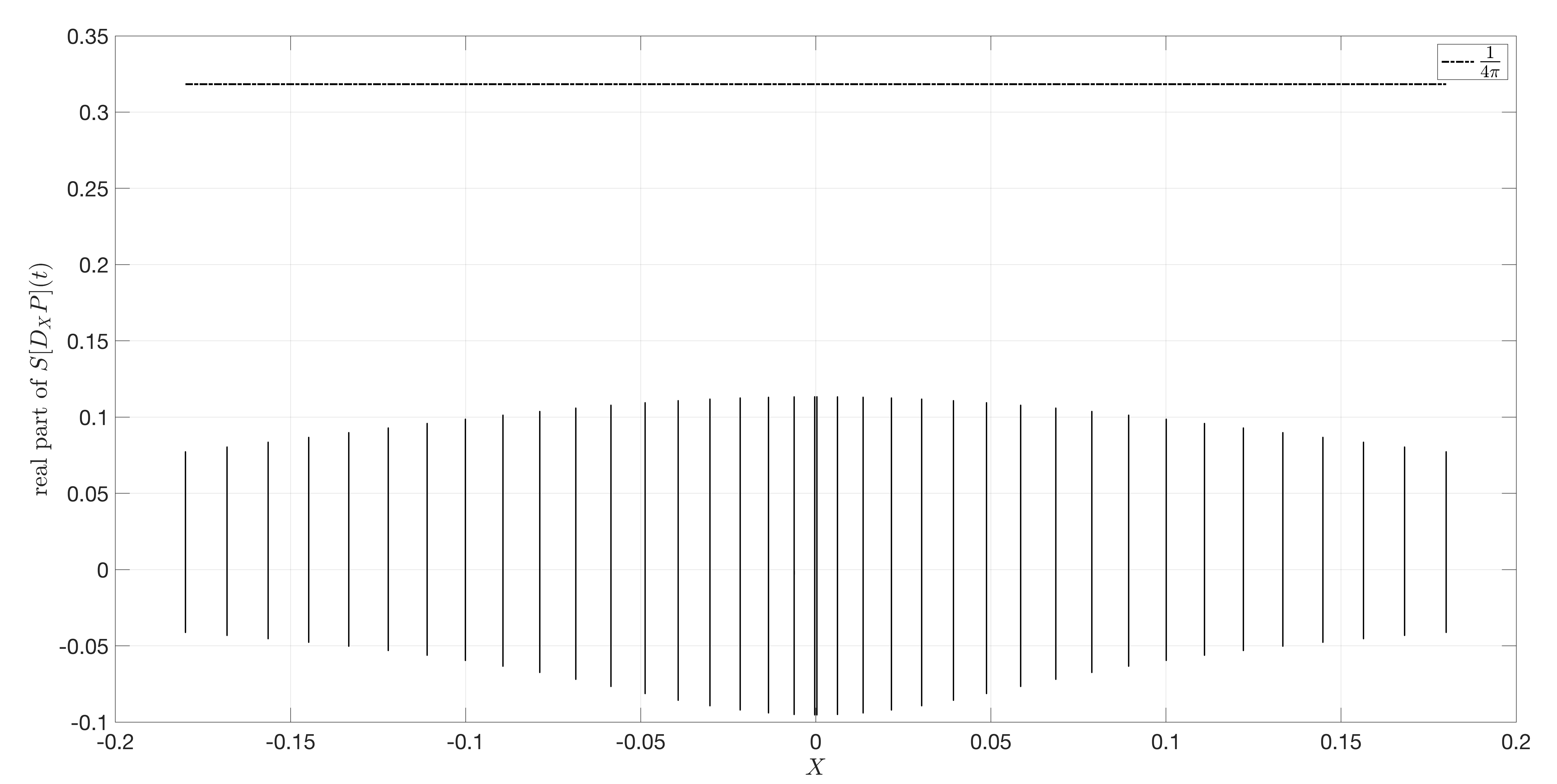}
\caption{
Numerical investigation of the stability condition for a stable  JONSWAP spectrum, cf. Section \ref{sec:conclusions} for more details. {We are using a target of $1/4\pi$ as in equation \eqref{eq:OdinALbCond}.}
{\bf Left:} Plots of the curve $\Gamma_X$ on the complex plane for different values of $X.$ Since $1/4\pi$ is always outside the $\Gamma_X,$ this spectrum is stable.
{\bf Right:} The span of the real parts of $\Gamma_X$ for different values of $X.$ 
}
\label{Fig:1} 
\end{center}
\end{figure}

\begin{figure}[h!]
\begin{center}
\includegraphics[width=0.33\textwidth]{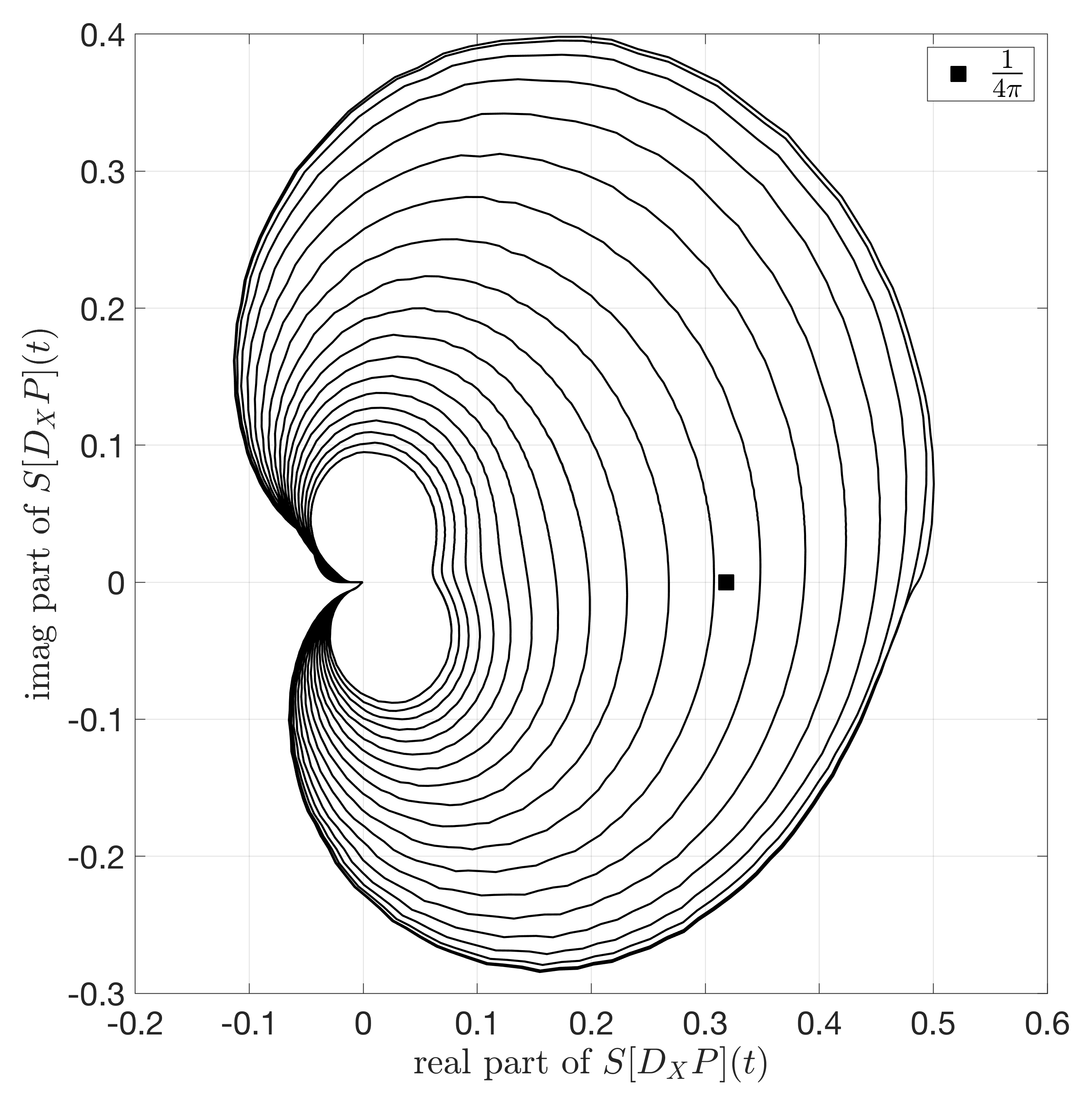} 
\includegraphics[width=0.66\textwidth]{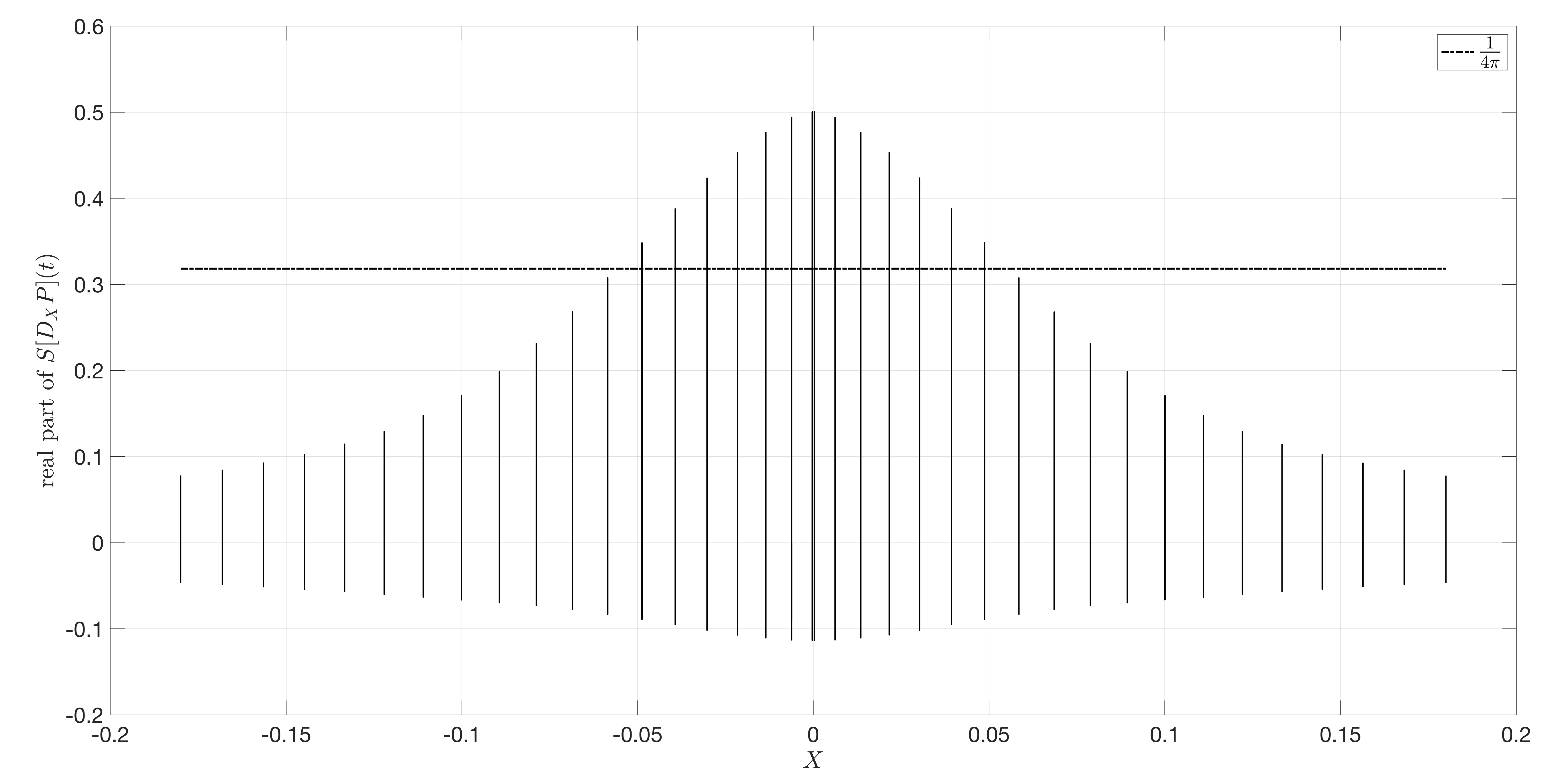}
\caption{
Numerical investigation of the stability condition for an unstable  JONSWAP spectrum. 
{\bf Left:} Plots of the curve $\Gamma_X$ on the complex plane for different values of $X.$  Since $1/4\pi$ is contained in some curves $\Gamma_X,$ the spectrum is unstable.
{\bf Right:} The span of the real parts of $\Gamma_X$ for different values of $X.$ In this case it highlights clearly the bandwidth of unstable wavenumbers $X.$ 
}
\label{Fig:3} 
\end{center}
\end{figure}

\subsection{1 versus 2 spatial dimensions}

It must be noted that in the original paper \cite{Alber1978} a two-dimensional setup is used, with the Davey-Stewartson equation for the envelope instead of the NLS equation \eqref{eq:NLS}. However, while technically two-dimensional, the Davey-Stewartson equation has  unidirectional propagation built in, and the second dimension is merely the ``transverse direction''.  This leads to Alber's ``eigenvalue relation'' eventually being one-dimensional: an effective spectrum is used, that results from appropriate integration of the two-dimensional spectrum along the transverse direction. In that sense, Theorem \ref{thrm:PenroseConditionResolve} can be  used in $1+1$ dimensional scenarios automatically, as the effective stability condition is one-dimensional anyway.

In genuinely two-dimensional settings (e.g. crossing seas), things are more complicated: the NLS equation \eqref{eq:NLS} is no longer an appropriate model. Systems of NLS equations \cite{Onorato2006,Shukla2006,Steer2019} or systems of other dispersive equations \cite{Gramstad2018} have been proposed. In any case the departure point is no longer  a single scalar NLS equation.

\subsection{Other problems}
More broadly, it must also be mentioned that combining NLS-type equations with stochastic modelling is natural in many different contexts, not only ocean waves.
It is thus natural that 
variants of the Alber equation are being independently rederived in different branches of physics, including optics \cite{Han2017} and many-particle systems \cite{Dubertrand2016}. Thus the main results of this paper are, in principle, applicable and/or generalisable to  other problems as well.

}

\begin{figure}[hp!]
\includegraphics[width=0.89\textwidth]{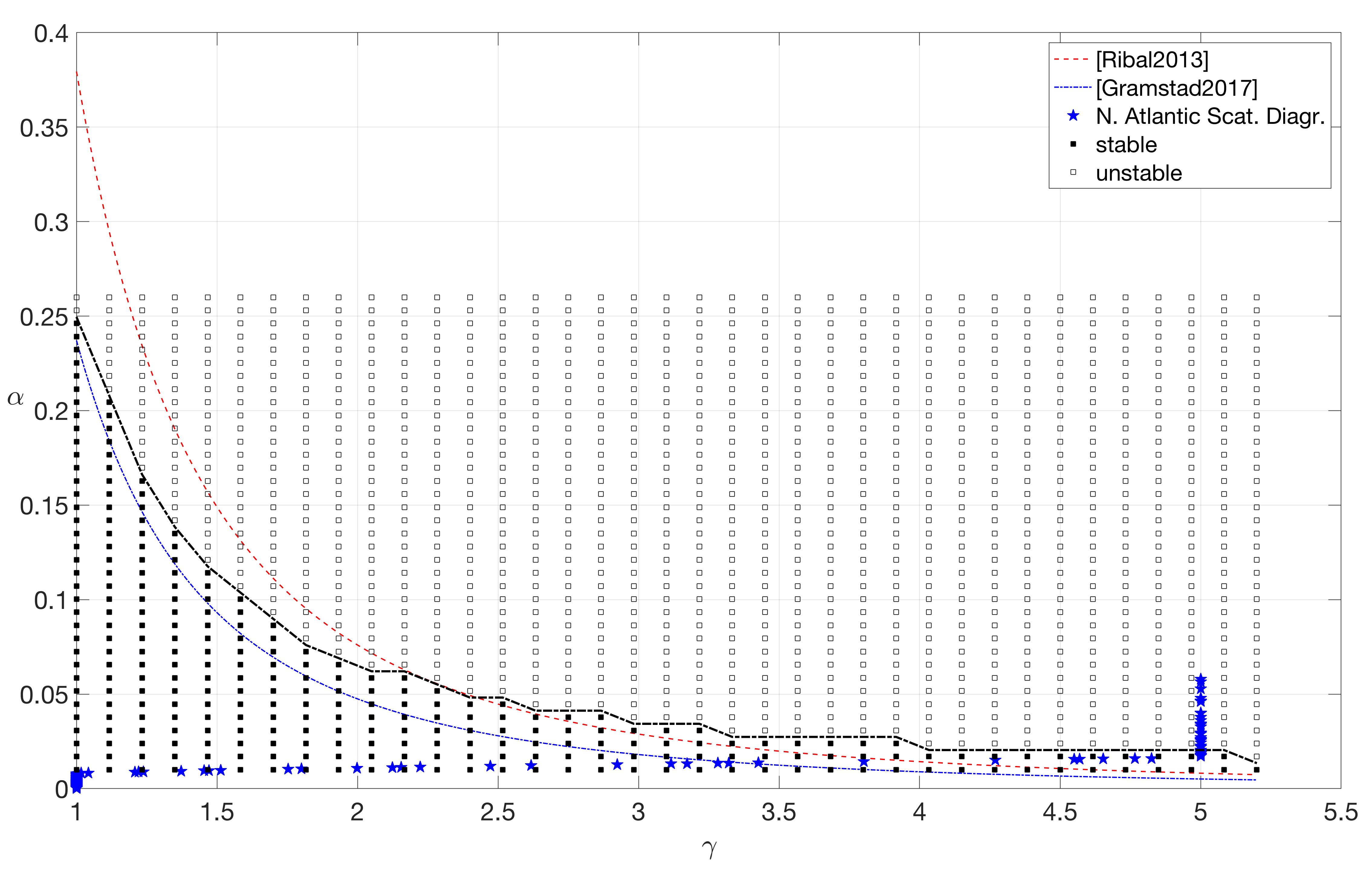} \\
\includegraphics[width=0.89\textwidth]{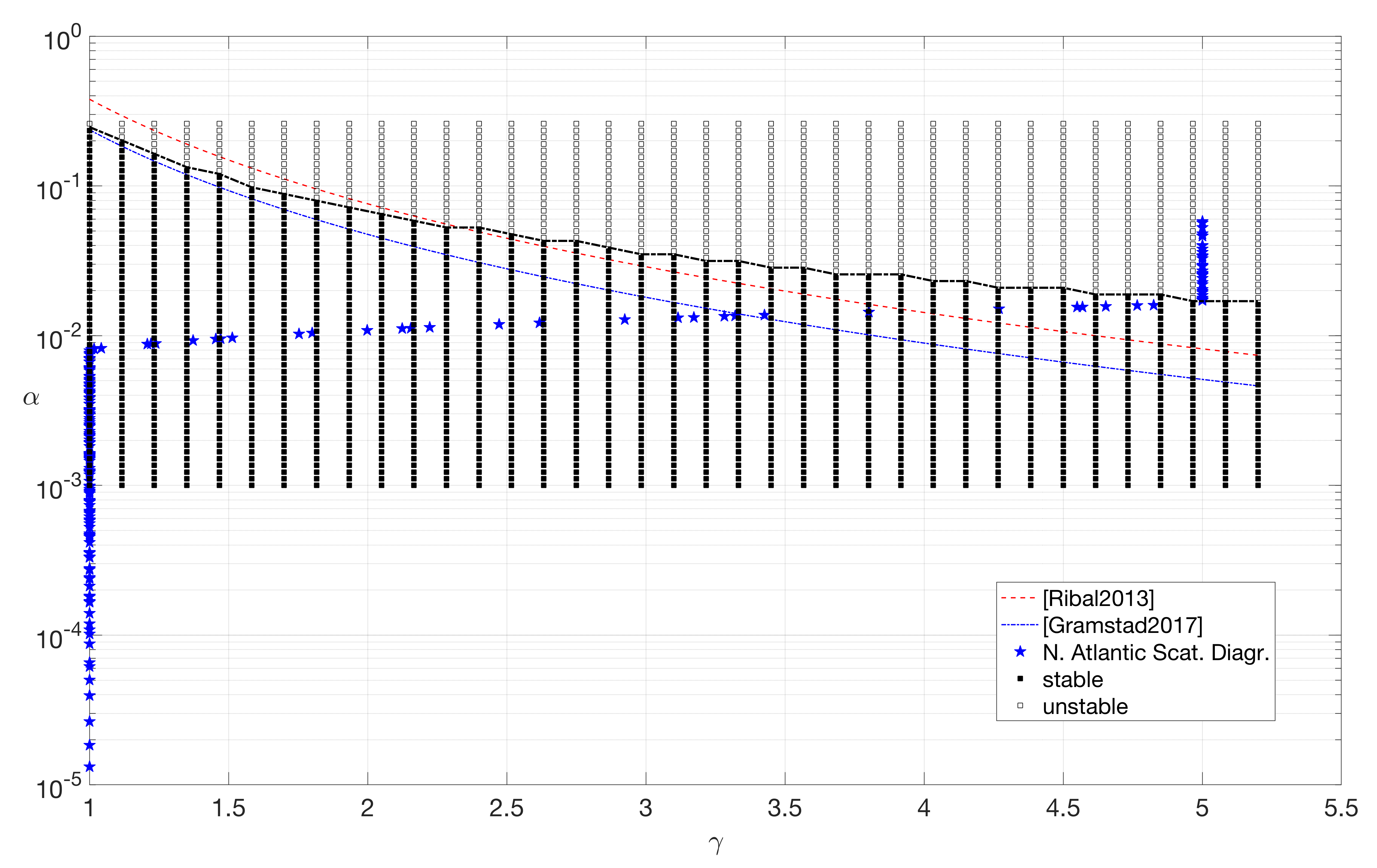}
\caption{
A number of points on the $(\gamma,\alpha)$ plane are tested for  stability of the corresponding JONSWAP spectrum, cf. equation \eqref{eq:defeqJONSWpsp}. $\alpha$ controls the power of the sea state (larger $\alpha$ means larger significant wave height) and $\gamma$ controls the effective bandwidth (larger $\gamma$ means more narrowly peaked spectrum). The carrier wavenumber $k_0$ can easily be seen not to affect the (in)stability of the spectrum.  $(\gamma,\alpha)$ points found to be stable are marked with a full square, while points found to be unstable are marked with an empty square. For reference the proposed separatrices of \cite{Ribal2013a} and \cite{Gramstad2017} are shown (they are of the form $\alpha\cdot\gamma/\beta =C,$ where $\beta$ is the mean wave steepness and $C=0.77$ \cite{Gramstad2017} or $C=0.974$ \cite{Ribal2013a}). More details can be found in Section \ref{sec:conclusions}. {\bf Top:} Linear scaling in both axes. {\bf Bottom:} Log scaling in the $\alpha$ (vertical) axis.
}
\label{Fig:synoptic} 
\end{figure}

\bigskip

\noindent {\bf Acknowledgment:} We would like to thank C. Saffirio, O. Gramstad and A. Babanin for helpful discussions on various aspects of this work.



\vskip2pc

\bibliography{notes_biblio_3.bib}
\bibliographystyle{siam}

 { }

\appendix

\section{Auxiliary lemmata}

\begin{lemma} \label{lm:myyoyngineq}
Let $A,B>0,$ $\zeta\in(0,1).$ Then
\[
\frac{1}{A+B} \leqslant \frac{1}{A^\zeta B^{1-\zeta}}
\]	
\end{lemma}
 \noindent {\bf Proof:} The well-known Young's inequality for products implies that,
for $a,b>0,$ $p\in(1,\infty),$ $\frac{1}{p}+\frac{1}q=1,$ 
\[
ab\leqslant \frac{a^p}{p} + \frac{b^q}{q} \leqslant a^p+b^q.
\]
Now setting $A=a^p,$ $B=b^q$ we have
\[
A^{1/p} B^{1/q}  \leqslant A+B \quad \Rightarrow \quad \frac{1}{A+B} \leqslant \frac{1}{A^{1/p} B^{1/q}}.
\]
By setting $\zeta=1/p$ and observing that $1/q=1-1/p=1-\zeta$ the conclusion follows.
\qed

\begin{lemma}\label{lm:decay_htilde}
Let $\widetilde{h}(X,s)$ be as in equation \eqref{eq:h_htilde}. Then
\[
\lim\limits_{\rho\to\infty}\sup\limits_{\substack{\real(\omega)>0\\ |\omega|>\rho}} |\widetilde{h}(X,\omega)|=0
\]
\end{lemma}

\noindent {\bf Proof:} Recall that $P\in\mathcal{S}(\R)$ is of compact support. Hence by construction $XD_XP(k)=P(k+\frac{X}2)-P(k-\frac{X}2)$ is also of compact support for each $X\in\R.$ Let $M=M(X)$ be such that $\mathop{supp} XD_XP\subseteq [-M,M].$ Then for all $\rho$ large enough we have
\[
\begin{aligned}
G(\rho)&:=
\sup\limits_{\substack{\real(\omega)>0\\ |\omega|>\rho}} |\widetilde{h}(X,\omega)|\leqslant \sup\limits_{\substack{\real(\omega)>0\\ |\omega|>\rho}} |q| \int_{\R} \frac{|XD_XP(k)|}{|\omega-4\pi^2 i pXk|} dk=
|q| \sup\limits_{\substack{\real(\omega)>0\\ |\omega|>\rho}}  \int_{-M}^M \frac{|XD_XP(k)|}{|\omega-4\pi^2 i pXk|} dk\\
&\leqslant 
|q|\int_{\R} {|XD_XP(k)|}dk \sup\limits_{\substack{\real(\omega)>0\\ |\omega|>\rho, \,\,|k|<M}}  \frac{1}{|\omega-4\pi^2 i pXk|}.
\end{aligned}
\]
Clearly $\lim\limits_{\rho\to\infty} G(\rho)=0.$
\qed

\begin{theorem}[Conditional integrability of the Hilbert transform] \label{thrm:l1hilbertzerointegral}
	Let $f\in\mathcal{S}(\R)$ be a function of compact support with $\int_tf(t)dt=0.$ Then
\[
\|\mathbb{H}[f]\|_{L^1(\R)}<\infty.
\]
\end{theorem}

\noindent {\bf Proof:} Choose an $M>0$ so that the support of $f$ is contained in $[-M,M],$ i.e.\  $f(x)=0$ $\forall |x|\geqslant M.$ We will also use the ``double'' interval,  $J:=[-2M,2M]$ and its complement $J^c=\R\setminus J.$ By an elementary estimate we have
\[
\|\mathbb{H}[f]\|_{L^1(\R)} \leqslant 4M\|\mathbb{H}[f]\|_{L^\infty(\R)}+ \int_{J^c} |H[f](x)|dx\leqslant 4CM\|\mathbb{H}[f]\|_{H^1(\R)}+ \int_{J^c} |H[f](x)|dx
\]
where $C$ is the constant of the Sobolev embedding $H^1(\R) \hookrightarrow L^\infty(\R)$. Moreover, using the fact that $\int_\R f(t)dt=0$ we have
\[
\begin{aligned}
I&:=\int_{J^c} |H[f](x)|dx =
\frac{1}\pi \int_{J^c}\Big |\int_{\R} \frac{f(t)}{x-t}dt \Big| dx=
\frac{1}\pi \int_{J^c} \Big|\int_{\R} \left( \frac{f(t)}{x-t} -\frac{f(t)}{x}\right)dt \Big|dx=\\
&=\frac{1}\pi \int_{J^c} \Big|\int_{\R} f(t)\left( \frac{1}{x-t} -\frac{1}{x}\right)dt\Big|dx
=\frac{1}\pi \int_{J^c} \Big|\int_{-M}^{M} f(t)\frac{t}{x(x-t)}dt \Big|dx,
\end{aligned}
\]
where in the last step we also used the fact that  $f$ is supported inside $[-M,M].$
Now observe that for any $x\notin[-2M,2M]$ and  $t\in[-M,M]$ 
\[
|t|\leqslant |x-t| \quad \Rightarrow \quad |x|=|x+t-t|\leqslant 2|x-t|\quad \Rightarrow \quad \frac{1}{|x-t|} \leqslant \frac{2}{|x|}. 
\]
Hence
\[
\left| \frac{t}{x(x-t)} \right|\leqslant \frac{2M}{x^2} \qquad \Rightarrow \qquad 
I\leqslant \frac{2M}{\pi} \int_{J^c} \frac{1}{x^2}dx \int_{-M}^M
|f(t)|dt <\infty.
\]
\qed


\section{Derivation of the Alber equation}

\label{sec:app__A}

{ 
\begin{remark}
Technically, the Alber equation does not govern any moments of solutions of NLS. It is derived heuristically, assuming the existence of a stochastic solution for the NLS with a certain kind of autocorrelation function and then applying a Gaussian closure to the resulting infinite moment hierarchy. So while, in certain situations, it may well turn out to be a reasonable approximation for certain second moments of solutions of the NLS equation, we don't study such an approximation in this paper.

In what follows in this Section we describe systematically the steps for the heuristic derivation of the Alber equation from the NLS equation. 
In particular, we use exact properties of certain Gaussian processes which are natural in the linear theory of water waves in order to motivate and justify the Gaussian closure used.

It is important to note that other equations of a similar character can be derived using different assumptions, cf. e.g. \cite{Andrade2018,Stuhlmeier2018}, and the results of this paper could motivate analogous advances for those equations as well.
\end{remark}
}

To explain the derivation of the Alber equation \eqref{eq:Alber1} as a second moment of the NLS \eqref{eq:NLS},  {first consider }  the algebraic (deterministic) second moment: denoting
\[
R_1(\alpha,\beta,t):=u(\alpha,t)\overline{u}(\beta,t),
\]
a straightforward computation leads to
\beq
i\partial_t R_1 + \frac{p}2 \left(\Delta_\alpha - \Delta_\beta\right) R_1
+\frac{q}2 R_1(\alpha,\beta,t) \left[ R_1(\alpha,\alpha) - R_1(\beta,\beta) \right]  =0
\ee
for the evolution in time of $R_1.$  {Thus,} despite taking a second moment of a nonlinear equation, the {\em exact algebraic moment closure}
\[
|u(\alpha,t)|^{2} u(\alpha,t)\overline{u}(\beta,t) = R_1(\alpha,\alpha,t)R_1(\alpha,\beta,t)
\]
allows one to have a closed, exact second moment equation. The same equation is called the ``infinite system of fermions'' in statistical physics \cite{Chen2015}. 
Now consider the stochastic second moment, 
\[
R(\alpha,\beta,t):=\Expe [u(\alpha,t)\overline{u}(\beta,t)].
\] 
Obviously now the algebraic closure is not enough, as
$
\Expe [|u(\alpha,t)|^{2} u(\alpha,t)\overline{u}(\beta,t)]
$
 is a fourth order stochastic moment, and not exactly expressible in terms of second order moments in general. However, for Gaussian processes (under additional assumptions described below) it can be seen that
\beq\label{eq:gmcl098}
\Expe[|u(\alpha,t)|^{2} u(\alpha,t)\overline{u}(\beta,t)] = 2 R(\alpha,\alpha,t)R(\alpha,\beta,t).
\ee
This is reminiscent of the well known real-valued Isserlis Theorem; the difference is that here $u$ is complex valued (and the factor $2$ is an artifact of the complex-valuedness of $u$). So the Alber equation \eqref{eq:Alber1} and the deterministic Wigner transform of the Schr\"odinger equation \eqref{eq:NLS} differ only in terms of this factor of $2.$

The precise result we invoke here can be summarised as follows:

\begin{observation}[A complex Isserlis theorem]\label{obs:ComplexIsserlis}
{A moment closure result is proved  in \cite{Reed1962}, and  a special case of it is the following:}

Let $z(x)$ be a Gaussian, zero-mean, stationary process with the additional property that
\beq\label{eq:circsymm1}
E[u(x)u(x')]=0 \qquad  \forall x,x'\in\R.
\ee
Then 
\[
E[\overline{z(x_1)z(x_2)}z(x_3)z(x_4)]=E[\overline{z(x_1)}z(x_3)]E[\overline{z(x_2)}z(x_4)]+E[\overline{z(x_2)}z(x_3)]E[\overline{z(x_1)}z(x_4)].
\]

This result directly implies the closure relation
\beq\label{eq:AlbMomCLo}
E[\overline{u(\alpha,t)u(\beta,t)}u(\alpha,t)u(\alpha,t)]=2E[\overline{u(\alpha,t)}u(\alpha,t)]E[\overline{u(\beta,t)}u(\alpha,t)],
\ee
which is exactly equation \eqref{eq:gmcl098}.

Moreover, the condition \eqref{eq:circsymm1} is equivalent to {\em circular symmetry}, i.e. to the condition that
\beq\label{eq:gaugeinv}
\{ e^{i\theta} u(x)\}_{\theta\in[0,2\pi)} \mbox{ are identically distributed for all } \theta \in[0,2\pi)
\ee
by virtue of a result by Grettenberg \cite{Wahlberg2005}.
\end{observation}

\begin{remark}[Physical meaning of the Gaussian closure]
Assuming that, for each $t_0$ the wave envelope $u(x,t_0)$ is a Gaussian process, with mean zero, stationary in $x$ (i.e. spatially homogeneous) and gauge invariant, $e^{i\theta} u(x,t_0) \sim u(x,t_0),$ is in line with standard modelling assumptions for linearised ocean waves \cite{Ochi1998}. 
In other words, the Gaussian moment closure of equation \eqref{eq:gmcl098} can be thought of as a linearisation of the probability structure of the wave envelope.
\end{remark}

By using the Gaussian closure \eqref{eq:gmcl098} we see that $R(\alpha,\beta,t)$ satisfies the equation
\beq
i\partial_t R + \frac{p}2 \left(\Delta_\alpha - \Delta_\beta\right) R
+q R(\alpha,\beta,t) \left[ R(\alpha,\alpha) - R(\beta,\beta) \right]  =0,
\ee
which is structurally the same as the infinite system of fermions, the only difference being an effective doubling of the coupling constant, $q/2\mapsto q.$
Introducing the assumption
\[
R(\alpha,\beta,t)=\Gamma(\alpha-\beta)+ \epsilon \rho(\alpha,\beta,t),
\]
we postulate that $R$ is in leading order homogeneous in space, and we set up an initial value problem for the inhomogeneity $\rho(\alpha,\beta,t),$
\beq \label{eq:linearisedF2rot00}
{ \bac
i\partial_t \rho + \frac{p}2 \left(\Delta_\alpha - \Delta_\beta\right) \rho
+q \left[ \Gamma(\alpha-\beta) + \epsilon \rho(\alpha,\beta) \right] \left[ \rho(\alpha,\alpha) - \rho(\beta,\beta) \right]  =0.
\ea}
\ee
Now denote
$\mathcal{R}$ be the rotation operator on phase-space
\beq
 \mathcal{R} [f(x,y)] := f(x+\frac{y}2,x-\frac{y}2), 
\ee
and consider the {\em average Wigner transform of the wave envelope} \cite{Athanassoulis2008,Athanassoulis2009}
\beq\label{eq:defaverageWT56}
\begin{aligned}
W(x,k,t) &= \int_{\R^d}  e^{-2\pi i k y} \Expe \big[u(x+\frac{y}2,t)\overline{u(x-\frac{y}2,t)}\big] dy =\mathcal{F}_{y\to k} \mathcal{R} [R(x,y,t)]=\\
&=\mathcal{F}_{y\to k} [\Gamma(y)+\epsilon\rho (x+\frac{y}2,x-\frac{y}2,t)]=P(k) + \epsilon w(x,k,t).
\end{aligned}
\ee
Then the Alber equation \eqref{eq:Alber1} is the equation for $w(x,k,t),$ i.e. it results by applying $\mathcal{F}_{y\to k} \mathcal{R}$ to equation \eqref{eq:linearisedF2rot00}.

{ 
So finally the relation between the unknown of the Alber equation, $w(x,k,t),$ and the wave envelope, $u(x,t),$ is 
\[
\mathcal{F}_{y\to k} \Expe [u(x+\frac{y}2,t)\overline{u}(x-\frac{y}2,t)] \approx P(k) + \epsilon w(x,k,t),
\]
where the quality of the approximation rests crucially on how accurate  the Gaussian closure is.}

Moreover, if $\int_{\R^{2d}}w_0(x,k)dxdk=0$ we have just an inhomogeneous redistribution of the energy of the homogeneous sea state, while if $\int_{\R^{2d}}w_0(x,k)dxdk>0$ we have a wave-train of finite energy interacting with a homogeneous sea state of infinite energy.

\section{Background results on Laplace and Hilbert transforms}

%
%

\begin{theorem}[Regularity of the Hilbert \& signal transforms] \label{thrm:HilbReg} Let $1<p<\infty.$ Then there exist constants $C=C(p)$ such that
\[
\|\mathbb{H}[u]\|_{L^p(\R)}\leqslant C \|u\|_{L^p(\R)}, \qquad \|\mathbb{S}[u]\|_{L^p(\R)}\leqslant (1+C) \|u\|_{L^p(\R)}.
\]
Moreover, $C(2)=1$ and for any $s\in \N,$
\[
\|\mathbb{H}[u]\|_{H^s(\R)} =\|u\|_{H^s(\R)}, \qquad \|\mathbb{S}[u]\|_{H^s(\R)} \leqslant 2\|u\|_{H^s(\R)}.
\]
Combining this with the Sobolev embedding $H^1(\R)\hookrightarrow C^0(\R)$ it follows that
\[
u\in H^1(\R) \quad \Rightarrow \quad \mathbb{H}[u], \mathbb{S}[u] \in C^0(\R).
\]
\end{theorem}

%

\begin{theorem}[Sokhotski-Plemelj formula]\label{thrm:SokhPlem}
For $u\in C(\R)\cap L^1(\R)$ and for any $s,c\in\R$
\[
\lim\limits_{\eta\to 0^+} \mathbb{H}[u]\Big(\frac{s- i \eta}{c}\Big) =  \mathbb{S}[u]\Big(\frac{s}c\Big).
\]
\end{theorem}

\begin{theorem}[Inverse Laplace transform, open half-plane] \label{thrm:InvLaplaceOPen}
Let $F(\omega)$ be a bounded analytic function on an open right half-plane, $\omega\in \Pi(M):= \{ \real z > M\}.$ Assume moreover that the limit $F_{M^+}(b):=\lim\limits_{\varepsilon\to 0^+} F(M+\varepsilon+ib)$ exists for all $b\in\R$ and is a continuous function in $b.$ 
Moreover assume that 
\beq\label{eq:L1constraint}
\lim\limits_{\rho\to +\infty} \sup\limits_{\substack{\omega\in \Pi(M) \\ |\omega|>\rho}} |F(\omega)| =0\qquad \mbox{ and } \qquad
\int_{-\infty}^{+\infty} |F_{M^+}(s)| ds < \infty.
\ee
Then
\[
F(\omega) = \mathcal{L}_{t\to \omega} [f(t)] \qquad \mbox{ where } \qquad f(t) = \frac{e^{Mt}}{2\pi} \int_{-\infty}^{+\infty} e^{ist} F_{M^+}(s) ds,
\]
i.e.
\[
\mathcal{L}^{-1}_{\omega\to t} [F] = \frac{e^{Mt}}{2\pi} \int_{-\infty}^{+\infty} e^{ist} F_{M^+}(s) ds.
\]
\end{theorem}

\section{Moments and Derivatives of the Alber-Fourier equation} \label{sec:appdeiff}

Denote
\beq \label{eq:defBilBrack22}
\begin{aligned}
& L[P_1\neg P_2; m] := \Big[P_1\Big(k-\frac{X}2\Big) - P_2\Big(k+\frac{X}2\Big)\Big] m(X,t)\\
& N[m;f_1\neg f_2] := \int_s m(s,t) \Big[f_1\Big(X-s,k-\frac{s}2,t\Big) - f_2\Big(X-s,k+\frac{s}2,t\Big)\Big] ds.
\end{aligned}
\ee
The nonlinearity $\mathbb{B}[m,f]$ defined in equation \eqref{eq:defBilBrack} is comprised of 
\[
\mathbb{B}[m,f] = iq L[P\neg P;m] + \epsilon i q N[m;f\neg f].
\]

\begin{lemma}  \label{lm:2aux12} For any multi-indices $\alpha,\beta,\gamma,\delta \in (\mathbb{N}\cup \{0\})^d$ we have the following relations 
\[
\begin{aligned} 
& X^\alpha L[P_1\neg  P_2;m] =  L[P_1\neg P_2; X^\alpha m], \\
& k^\beta L[P_1\neg P_2;m] =  \sum\limits_{0\leqslant \beta' \leqslant \beta}\binom{\beta}{\beta'}L[k^{\beta-\beta'} P_1 \neg (-1)^{\beta'} k^{\beta-\beta'} P_2; (\frac{X}2)^{\beta'} m],
\\
& \partial_X^\gamma L[P_1\neg P_2;m] = \sum\limits_{0\leqslant \gamma'\leqslant \gamma} \binom{\gamma}{\gamma'} 
L[(-1/2)^{\gamma-\gamma'} \nabla^{\gamma-\gamma'}P_1 \neg (1/2)^{\gamma-\gamma'} \nabla^{\gamma-\gamma'} P_2; \partial_X^{\gamma'}m],\\
& \partial_k^\delta L[P_1\neg P_2;m] = L[\nabla^\delta P_1\neg \nabla^\delta P_2;m],
\end{aligned} 
\]
and 
\[
\begin{aligned}
& X^\alpha N[m;f_1\neg f_2] =  \sum\limits_{0\leqslant \alpha' \leqslant \alpha} \binom{\alpha}{\alpha'}N[X^{\alpha-\alpha'}m;X^{\alpha'}f_1\neg X^{\alpha'}f_2], \\
& k^\beta N[m;f_1\neg f_2] =  \sum\limits_{0\leqslant \beta' \leqslant \beta} \binom{\beta}{\beta'}N\Big[{\Big(\frac{X}2\Big)^{\beta-\beta'}m};k^{\beta'}f_1\neg (-1)^{\beta-\beta'}k^{\beta'}f_2\Big], 
\\
& \partial_X^\gamma N[m;f_1\neg f_2] = 
 N[m;\partial_X^\gamma f_1\neg \partial_X^\gamma f_2],
\\
& \partial_k^\delta N[m;f_1\neg f_2] =  N[m;\partial_k^\delta f_1\neg \partial_k^\delta f_2].
\end{aligned} 
\]
Moreover,
\[
\begin{aligned} 
X^\alpha k^\beta \partial_X^\gamma\partial_k^\delta \big( k \cdot X f \big) =
k \cdot X \big( X^\alpha k^\beta \partial_X^\gamma\partial_k^\delta   f \big)  +X^\alpha k^\beta 
\sum\limits_{\substack{0\leqslant \gamma'<\gamma\\ 0\leqslant\delta'<\delta}} 
 \binom{\gamma}{\gamma'} \binom{\delta}{\delta'} \big( \partial_X^{\gamma'}\partial_k^{\delta'}   f \big)  \big( \partial_X^{\gamma-\gamma'}\partial_k^{\delta-\delta'} X\cdot k \big).
\end{aligned} 
\]
\end{lemma}
The proof follows from direct computations using the
definition of $L[P_1\neg  P_2;m] $ and $ N[m;f_1\neg f_2].$

\medskip 

By applying the operator $X^\alpha k^\beta \partial_X^\gamma \partial_k^\delta$ to equation \eqref{eq:wf1} and commuting according to Lemma \ref{lm:2aux12} one obtains equation \eqref{eq:aux123}
with right hand side
\[
\begin{aligned}
\mathbb{B}^{({\alpha,\beta,\gamma,\delta})}[f]& =  
-\sum\limits_{\substack{0\leqslant\gamma'\leqslant \gamma\\0\leqslant \delta'\leqslant\delta\\ |\gamma'+\delta'|<|\gamma+\delta|}} 
 \binom{\gamma}{\gamma'} \binom{\delta}{\delta'}  \big( \partial_X^{\gamma-\gamma'}\partial_k^{\delta-\delta'} X\cdot k \big)  f^{\alpha,\beta,\gamma',\delta'}\\
 - q i 
\sum\limits_{\substack{0\leqslant \beta'\leqslant \beta \\ 0\leqslant \gamma'\leqslant \gamma}} &
(\frac{1}2)^{|\gamma-\gamma'|+|\beta'|} \binom{\beta}{\beta'}  \binom{\gamma}{\gamma'} L [k^{\beta-\beta'} (-1)^{|\gamma-\gamma'|}\nabla^{\gamma-\gamma'+\delta}  P \,\neg\, k^{\beta-\beta'} \nabla^{\gamma-\gamma'+\delta} P\,; \,\int_k f^{\alpha+\beta',0,\gamma',0} dk] \\
 -\epsilon q i   
\sum\limits_{\substack{0\leqslant \alpha'\leqslant \alpha \\ 0\leqslant \beta'\leqslant \beta}} &
(\frac{1}2)^{|\beta-\beta'|}\binom{\alpha}{\alpha'} \binom{\beta}{\beta'} N[\int_k f^{\alpha-\alpha'+\beta-\beta',0,0,0} dk \,;\,  
f^{\alpha',\beta',\gamma,\delta} \,\neg\, (-1)^{|\beta-\beta'|} f^{\alpha',\beta',\gamma,\delta}].
\end{aligned}
\]

\end{document}